%% file: sample-acmlarge.tex
\documentclass[acmlarge]{acmart}

\usepackage{lscape}
\usepackage{array}
\usepackage[utf8]{inputenc}
\usepackage{graphicx}
\usepackage{longtable} 
\usepackage{tikz} 
\usepackage{lipsum} 

\usepackage[autostyle]{csquotes}
\AtBeginDocument{%
  \providecommand\BibTeX{{%
    \normalfont B\kern-0.5em{\scshape i\kern-0.25em b}\kern-0.8em\TeX}}}

\setcopyright{acmcopyright}
\copyrightyear{2025}
\acmYear{2025}
\acmDOI{XXXXXXX.XXXXXXX}

\acmJournal{POMACS}
\acmVolume{XX}
\acmNumber{X}
\acmArticle{XXX}
\acmMonth{5}
\begin{document}

\title {Review-based Recommender Systems: A Survey of Approaches, Challenges and Future Perspectives}

\author{Emrul Hasan}
\authornote{\textbf{Equal contribution. Sorted by last name.}}
\email{e1hasan@torontomu.ca}
\affiliation{%
  \institution{Toronto Metropolitan University \& Vector Institute}
  \streetaddress{350 Victoria Street}
  \city{Toronto}
  \country{Canada}
}
\author{Mizanur Rahman}
\authornotemark[1]
\affiliation{%
  \institution{York University \& Royal Bank of Canada}
  \city{Toronto}
  \country{Canada}}
\email{mizanurr@yorku.ca}

\author{Chen Ding}
\affiliation{%
  \institution{Toronto Metropolitan University}
  \streetaddress{350 Victoria Street}
  \city{Toronto}
  \country{Canada}}
\email{cding@torontomu.ca}

\author{Jimmy Xiangji Huang}
\affiliation{%
  \institution{York University}
  \city{Toronto}
  \country{Canada}}
\email{jhuang@yorku.ca}

\author{Shaina Raza}
\affiliation{%
  \institution{Vector Institute}
  \streetaddress{108 College St., Suite W1140}
  \city{Toronto}
  \country{Canada}}
\email{Shaina.raza@vectorinstitute.ai}

\renewcommand{\shortauthors}{Hasan, Rahman, et al.}

\begin{abstract}
Recommender systems play a pivotal role in helping users navigate a vast selection of products and services. On online platforms, users have the opportunity to share feedback in various modes, such as numerical ratings, textual reviews, and likes/dislikes. Traditional recommendation systems rely on users’ explicit ratings or implicit interactions (e.g., likes, clicks, shares, and saves) to learn user preferences and item characteristics. Beyond numerical ratings, textual reviews provide insights into users’ fine-grained preferences and item features. Analyzing these reviews is crucial for enhancing the performance and explainability of personalized recommendation results. In this paper, we provide a comprehensive overview of the development in review-based recommender systems over recent years, highlighting the importance of reviews in recommender systems, as well as the challenges associated with extracting features from reviews and integrating them into ratings. Specifically, we introduce a classification scheme in terms of both the integration of reviews into recommendation systems and the technical methodology. Additionally, we summarize the state-of-the-art methods, analyzing their unique features, effectiveness, and limitations. The study also presents the various evaluation metrics, comparative analysis, datasets, and real-world applications of review-based recommendation systems. Finally, we propose potential directions for future research, including multi-modal data integration, multi-criteria rating information, and ethical considerations.
\end{abstract}




\maketitle
\input{Introduction}
\input{Background}
\input{ReviewBasedRec}

\input{Baselines}

\input{Datasets}

\input{EvaluationMetrics}
\input{Challenge}

\input{Discussion}

\section{CONCLUSIONS AND FUTURE WORK}
This paper presents a comprehensive exploration of review-based recommender systems. We classify these systems using two frameworks: (a) review utilization strategies (generic review methods, aspect-based methods, rating-review fusion, and ratings and aspects fusion) and (b) methodologies (probabilistic models, deep learning, and miscellaneous models). Through a detailed analysis of state-of-the-art methodologies, we show how leveraging user-generated reviews can enhance the personalization and accuracy of recommendations, surpassing the limitations of traditional rating-based systems. We identify the key challenges, including the complexity of learning user/item representations from reviews, review integration strategies, the need for scalable solutions to handle vast datasets, and the importance of ethical considerations in data handling. Despite these challenges, advancements in machine learning, deep learning, and NLP techniques offer promising avenues for extracting valuable insights from reviews, thereby enriching recommender systems with detailed understandings of user preferences. 

Looking forward, the integration of multi-modal data presents an exciting opportunity for further enriching review-based recommender systems, enabling a holistic approach to understanding user sentiments and preferences. Addressing data sparsity and the cold start problem remains a critical area for future research. Moreover, as recommender systems increasingly influence user choices and experiences, ethical considerations around bias mitigation and data privacy must be central to future developments. Ensuring these systems are fair, transparent, and privacy-compliant will be paramount in fostering long-term trust and acceptance among users.

\input{Acknowledgement}

\bibliographystyle{ACM-Reference-Format}
\renewcommand{\bibfont}{\fontsize{7pt}{6pt}\selectfont} 
\bibliography{sample-base}

\end{document}

%% file: Introduction.tex
\section{Introduction}

In the rapidly advancing digital era, the sheer volume of online content, including products, movies, restaurants, and services, presents a significant challenge: information overload. This overwhelming amount of information often makes it difficult for people to find what they truly need or prefer. In response, recommender systems have emerged as key solutions, guiding users to content that aligns with their interests and preferences. These systems play an important role in providing personalized recommendations, enhancing user experiences, contributing to business growth, and facilitating the decision-making process \cite{huang2020personalized}. Big technology companies including Google, Meta, Netflix and Amazon, extensively use these systems across a wide range of web domains such as e-commerce \cite{schafer1999recommender, hwangbo2018recommendation}, healthcare  \cite{tran2021recommender}, education \cite{tan2008learning}, news portals \cite{raza2019news}, and media streaming \cite{majid2013context}. 

Traditional approaches to recommender systems such as Collaborative Filtering (CF) \cite{su2009survey} and Content-based Filtering (CBF) \cite{thorat2015survey} have been used for developing personalized recommender systems. CF performs well when there is ample user rating data, but it struggles with sparse data, leading to inaccuracies and a lack of transparency \cite{xu2018adjective}. CBF addresses some limitations of CF by representing users and items through item attributes and user preferences \cite{mooney2000content}, but they also face challenges due to the dynamic nature of user interests \cite{thorat2015survey}. 

The rise of user reviews on online platforms has introduced a new dimension to recommendation systems. Unlike traditional systems that primarily rely on numerical ratings or browsing history, review-based systems leverage rich textual feedback, covering various elements such as product aspects, sentiment, and even writing style \cite{dang2021approach, hernandez2019comparative,artemenko2020using,al2023trusted,hazar2022learner}. Aspect-based opinion mining further enhances these systems by extracting detailed product attributes from reviews and aligning them with user preferences for more tailored recommendations \cite{ bhattacharya2022recent, hernandez2019comparative}.  Additionally, sentiment analysis captures user attitudes and emotions, providing a deeper understanding of preferences \cite{dang2021approach,bhattacharya2022recent,musto2017multi}. These systems incorporate factors such as review structure, content richness, and contextual information to address the limitations of rating-based methods  \cite{wu2019context,liu2019daml,zheng2017joint, chen2015recommender,seo2017interpretable,chin2018anr,huang2020personalized, al2019multi,osman2021integrating}.  For example, in e-tourism, reviews can systematically be summarized for user preferences, leading to more accurate and targeted mobile recommendations \cite{artemenko2020using}.  

Review-based recommender systems incorporate user-generated reviews to offer more personalized, accurate, and explainable recommendations \cite{rokach2022recommender,wu2019context}. However, incorporating textual data into recommendation models presents unique challenges, such as extracting fine-grained user preferences and item characteristics, handling diverse and abundant textual data, and addressing concerns around bias and privacy \cite{raza2024nbias}. Furthermore, systems must now account for cross-domain applications and multi-language challenges, where multilingual reviews provide an additional complexity that recent models have begun to address using multilingual aspect-based sentiment analysis \cite{liu2021multilingual}. As a result, review-based recommender systems require advanced methodologies and innovative solutions to tackle these evolving challenges.

Recently, hybrid models that combine CF with sentiment analysis and aspect-based methods have emerged as key developments. These approaches aim to overcome limitations of the traditional models by incorporating both structured and unstructured data \cite{ziani2017recommender,hung2020integrating, osman2018sentiment, mishra2020sentiment,contratres2018sentiment,asani2021restaurant}. Utilizing natural language processing, and deep learning techniques into review-based systems has driven the development of sophisticated recommendation models \cite{shoja2019customer,preethi2017application,hu2020reviewer,chiranjeevi2023lightweight,sharma2024sentiment}. Multi-criteria systems, which consider different aspects such as service, price, and quality, have shown to improve both recommendation accuracy and user satisfaction \cite{musto2017multi,singh2024sentiment}. Recent studies have demonstrated how aspect-based sentiment analysis, combined with CF, can provide more nuanced recommendations, particularly in large-scale datasets \cite{ozcan2022deep}. Beyond aspects, elements such as sentiment polarity, opinion diversity, and even narrative structure contribute to a holistic understanding of user preferences.

While extensive research has reviewed traditional recommender systems  \cite{li2023recent,lu2015recommender,adomavicius2005toward, konstan2012recommender, burke2002hybrid, shi2014collaborative, jannach2015recommenders, khan2017cross, batmaz2019review}, deep learning-based systems \cite{zhang2019deep}, graph-based approaches \cite{gao2023survey, wu2022graph,gao2020deep}, multi-criteria methods \cite{al2019multi,musto2017multi}, session-based systems \cite{wang2021survey,huang2021graph}, context-aware recommender systems \cite{haruna2017context, raza2019progress}, knowledge-based \cite{bobadilla2013recommender} and reinforcement learning systems \cite{afsar2022reinforcement,wang2020kerl}, there remains a notable gap in surveys specifically focusing on review-based recommender systems. Although review-based recommender systems have been partially explored in earlier works \cite{chen2015recommender,al2019multi}, these studies are now quite outdated. For instance, Chen et al. \cite{chen2015recommender} conducted a survey highlighting the importance of creating user and item profiles by leveraging user-generated reviews. However, this survey is limited to research works utilizing reviews for user profile construction and multi-faceted opinion mining. Additionally, being published in 2015, the survey does not reflect recent progress in deep learning and NLP techniques (e.g., attention mechanism \cite{DBLP:journals/tkde/TalLHYA21}, transformer models \cite{vaswani2017attention}, LLMs \cite{li2023prompt}, Contrastive Learning \cite{xia2022hypergraph}, Reinforcement Learning \cite{wang2020kerl}, Graph Contrastive Learning \cite{you2020graph}, Counterfactual arguments \cite{tan2021counterfactual}, etc.) to learn users' fine-grained preferences and item attributes. Similarly, a survey on multi-criteria review-based recommendations was conducted by Al-Ghuribi and Noah \cite{al2019multi} in 2019, leveraging user-generated reviews to improve the performance of multi-criteria recommendation systems. Although this survey explores fine-grained sentiment analysis, it ignores several critical aspects of review-based recommendation systems, including the integration techniques of reviews with ratings, sentiment polarity at different levels, comprehensive user and item feature learning, and advanced representation learning techniques.  In 2021, Shalom et al. \cite{shalom2021natural} surveyed the application of NLP in recommendation systems. Their survey primarily focused on the use of reviews and explanation generation, while ignoring the detailed coverage of multifaceted representation methods and advanced feature learning techniques. Moreover, it does not focus on the latest advancements in review-based recommendation systems.

This paper bridges these gaps by offering a comprehensive review that introduces review integration strategies, analyzes the latest feature learning and deep representation learning methods, investigates domain-specific challenges, and suggests future research directions. By covering both aspect-based and broader review-aware techniques, this survey provides a holistic perspective on this rapidly evolving field \cite{margaris2020makes}. The key contributions of this survey can be summarized as follows:

\begin{enumerate}
    \item We systematically explore review-based recommender systems and introduce two classification schemes—one based on how reviews are incorporated into recommendation systems and the other on technical methodologies for learning review features. To the best of our knowledge, this is the first comprehensive survey focused on this domain. 
   \item We provide a comprehensive overview of the state-of-the-art models, analyzing unique features, effectiveness, limitations, datasets, evaluation methods, and real-world applications of recommendation systems.
   \item We address key challenges and recent developments, proposing research directions to tackle theoretical and practical gaps, focusing on developing robust and transparent systems.

\end{enumerate}

\setlength{\abovedisplayskip}{1pt} 
\textbf{Methodology:} To conduct a systematic review on review-based recommender systems, we collected the relevant papers from a wide range of sources, including proceedings of the prestigious conferences in this field (e.g., RecSys, SIGIR, CIKM, AAAI, WWW, WSDM, KDD, UMAP, EMNLP, ACL), Google Scholar, and IEEE Xplore. Our review period spans January 2015 to October 2024. We used specific search keywords such as \enquote{review-based recommender systems}, \enquote{review-aware recommender systems}, \enquote{aspect-based recommender systems}, \enquote{aspect-aware recommender systems}, and \enquote{sentiment analysis in recommender systems}. 
 
 For conference proceedings, we initially recorded a total number of papers that contain any of the above-mentioned keywords in their titles, resulting in a total of 14,862 papers across all conferences. For Google Scholar and IEEE Xplore, we utilized a different strategy, applying Boolean operators. Our search queries were structured as ( \enquote{review-based} OR \enquote{review-aware} OR \enquote{aspect-based} OR \enquote{aspect-aware} OR \enquote{user reviews} OR \enquote{user review elements} OR \enquote{sentiment analysis} ) AND \enquote{recommender systems}. The initial Google Scholar search returned a total of 17,200 papers, whereas IEEE Xplore reported a significantly lower count of 855 papers for a similar keyword search. After the initial search, we skimmed through the titles and created a preliminary shortlist consisting of 529 papers. In the next step, we performed a screening based on titles and abstracts, focusing on relevance to the goals of our study and ensuring a careful selection process. This screening involved two reviewers to enhance reliability, with discrepancies resolved through discussion, ultimately shortlisting 113 papers for detailed analyses from all sources.

\begin{figure}[ht]
\centering
\includegraphics[width=1.0\textwidth]{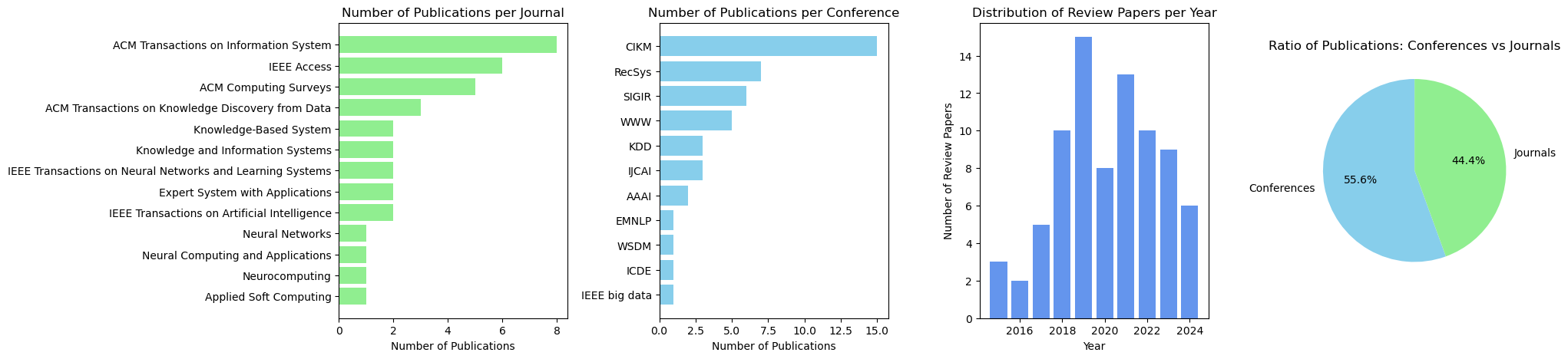}
 \Description{Trends in Review-based Recommender Systems (2015-–2024): Key publications in journals and conferences}
\caption{Trends in Review-based Recommender Systems (2015-–2024): Key publications in journals and conferences}
\label{fig:figure1}

\end{figure}

The selected papers undergo a thorough content analysis to extract information on methodologies, findings, and the evolution of trends over time. Figure \ref{fig:figure1} illustrates the distribution of the selected publications across different sources and formats, as well as their annual distribution.
 
To ensure the inclusion of the most relevant and impactful research, we applied the following inclusion criteria: (1) The paper must explicitly focus on review-based, review-aware, or aspect-based recommender systems. (2) The paper should introduce a novel methodology or present a significant enhancement to existing methods. (3) The paper must be peer-reviewed and published in a recognized NLP, AI or machine learning conference, such as those listed earlier. (4) Additionally, papers that explore the intersection of sentiment analysis and recommender systems, especially those utilizing user reviews to improve recommendation quality, were included. (5) For recent papers on arXiv, we required that they have at least 5 citations per year to ensure their influence in the field. (6) Lastly, we included several highly cited papers even if they were not from widely recognized venues.

For exclusion, we applied the following criteria: (1) Papers that focus solely on traditional collaborative filtering or content-based filtering without integrating user reviews or textual data were excluded. (2) Papers that dealt exclusively with non-recommender system applications, even if they involved sentiment analysis or aspect extraction, were also excluded. By following these criteria, we aimed to focus on research that was most relevant to the evolution of review-based recommender systems, ensuring that the selected papers would contribute meaningfully to our survey of the field.

%% file: Background.tex
\section{Background}

Recommender systems are broadly classified into three categories: Collaborative Filtering \cite{schafer2007collaborative}, Content-based Filtering \cite{pazzani2007content}, and Hybrid Systems \cite{thorat2015survey}. Before further exploring the review-based recommender systems, we will discuss each of these categories in detail in the following subsections.

\subsection{Collaborative Filtering}
Collaborative Filtering (CF) \cite{schafer2007collaborative} is a method used in recommender systems to predict a user's interests based on the preferences of a larger user group. This technique relies on collecting and analyzing data on user behaviors, activities, or preferences to make personalized recommendations \cite{su2009survey}. The core idea behind CF is that, if two users have shown similar preferences in their past interactions with items, they are likely to continue showing similar patterns in the future. CF establishes a relationship between two entities: users and items. 

CF can be categorized into two main types: Memory-based \cite{deshpande2004item, pu2011user}, and Model-based approaches \cite{paterek2007improving,sarwar2001item}. Memory-based approaches make predictions based on similarities between users or items, directly using past user interaction dat. Model-based approaches, on the other hand, involve building and training models from user-item interaction data to predict a user's interest in an item based on learned patterns. In CF systems, prediction is primarily based on historical data of user-item interactions. Memory-based methods make predictions by averaging the ratings of similar items or users, while model-based approaches predict user preferences by learning latent factors of users and items, often using techniques like matrix factorization \cite{kalman1996singularly}. 

Matrix factorization maps both the items and users by vectors of factors inferred from item ratings. To illustrate the matrix factorization technique, consider a user $u$ with associated vector $\mathbf{p}_u \in \mathbb{R}^d$ and an item $i$ with associated vector $\mathbf{q}_i \in \mathbb{R}^d$, where $d$ is the dimension of the vector. The dot product of these vectors represents the interaction between user and item, estimating the user’s overall preference toward the item, as follows:

\begin{equation}
    \hat{r}_{ui} = \mathbf{p}_u^T \mathbf{q}_i +b
\end{equation}
where b is the bias term. The core idea behind matrix factorization is to minimize the difference between the predicted rating score $\hat{r}_{ui}$ and the true rating score $r_{ui}$. This is done by solving the following objective function:

\begin{equation}
    \min_{P, Q} \sum_{(u,i) \in R} (r_{ui} - \mathbf{p}_u^T \mathbf{q}_i)^2 + \lambda (\|P\|^2 + \|Q\|^2)
\end{equation}
where $P$ is an $m \times k$ matrix representing user features with $k$ number of latent factors, $Q$ is a $k \times n$ matrix representing item features, and $\lambda \left( \|P\|^2 + \|Q\|^2 \right)$ is the regularization term
with $\lambda$ representing regularization parameter. The ultimate objective of the matrix factorization is to reconstruct the original matrix $R$ by the product of two smaller matrices $P \times Q$ as closely as possible.

\subsection{Content-based Filtering}
Content-based filtering (CBF) recommends items by analyzing their attributes and comparing them to a user's preferences or past interactions \cite{pazzani2007content}. This technique creates user profiles based on demographic information or previous interactions, then aligns these with item attributes for personalized recommendations. It addresses the cold start problem effectively by suggesting items based on their features, even without prior user interaction history.
While adept at recommending items similar to those a user has previously shown interest in, CBF can sometimes lack diversity in its suggestions. To overcome this, recent advancements leverage machine learning and natural language processing to refine user profiles and item descriptions, aiming to improve the accuracy, relevancy and diversity of recommendations \cite{yang2021aspect, wang2021leveraging, lyu2021reliable}. Despite its challenges, CBF remains a key method for personalized recommendations, with continuous enhancements driven by technological progress.

\subsection{Hybrid Recommender Systems}
Hybrid recommender systems combine elements from both collaborative filtering and content-based filtering to offer more accurate and diverse recommendations \cite{lucas2013hybrid}. By integrating these approaches, hybrid systems can leverage the strengths and mitigate the weaknesses of each method. They use a variety of techniques, including combining separate collaborative and content-based models \cite{de2010combining,choi2010hybrid}, incorporating features of one method into the other \cite{bedi2013modeling,mooney2000content}, or building a unified model that simultaneously learns user and item representations from both user interactions and item attributes  \cite{ccano2017hybrid}.
These techniques allow hybrid systems to provide personalized recommendations like content-based methods while also benefiting from the broad recommendation capabilities of collaborative filtering \cite{choi2010hybrid}. Hybrid systems are particularly effective in addressing the cold start problem for new users and items, as they can recommend items based on content when user interaction data is sparse \cite{thorat2015survey}.
For instance, sentiment analysis and multi-criteria review mining can extract valuable insights from user reviews, including specific opinions on product aspects and overall user sentiment. Sentiment-Aware Deep Learning \cite{hung2020integrating} and Aspect-Based Sentiment Analysis \cite{musto2017multi} have shown to enhance recommendation quality, particularly in scenarios with sparse user ratings or new items. However, implementing hybrid recommender systems can be complex and resource-intensive \cite{su2009survey}.

\section{Why adding review into recommender systems}
The ultimate goal of a recommender system is to suggest items that match user's interests. Traditional recommender systems learn user's preferences based on historical rating data. However, these rating-based systems encounter significant challenges, including data sparsity, lack of explainability, and  reduced accuracy. Incorporating additional forms of information, such as text and images, is advantageous when available. Reviews, in particular, play a crucial role in understanding user preferences and item characteristics.  User reviews, being free-form text, allow users to express their opinions in any direction. Utilizing reviews helps addressing these critical issues in recommender systems. Following are some of the key avantages of incorporating reviews.

\textbf{Addressing Data Sparsity: }
Data sparsity poses a significant challenge in recommender systems, particularly in real-world scenarios where rating data is scarce. Learning user preferences and item features from such sparse data is challenging. Reviews help address this challenge by providing detailed textual feedback, which is valuable for extracting features to understand user preferences. Various existing recommender systems demonstrate the effectiveness of leveraging reviews in mitigating data sparsity issues \cite{zheng2017joint, tan2016rating}.

\textbf{Improving Accuracy: }
Improving accuracy is another challenging task for any recommender system. Recommendation accuracy largely depends on accurately learning user and item representation. Since reviews contain fine-grained opinions, utilizing them for feature learning can be beneficial. Significant research has utilized reviews to improve accuracy \cite{kim2016convolutional, zhang2017joint,tan2008learning, seo2017interpretable}. Different NLP techniques are used to extract the features from user reviews and item reviews, followed by an interaction technique to predict the ratings \cite{chin2018anr,guan2019attentive}. For example, Chin et al. \cite{chin2018anr} utilize Word2Vec \cite{church2017word2vec} word embedding for vector representation of words for feature extraction and attention mechanism for aspect important estimation. Similarly, Guan et al. \cite{guan2019attentive} use a mutual attention mechanism to capture the fine-grained interaction information by establishing correlations between user and item contextual features across multiple feature subspaces.  Some studies utilize both reviews and ratings for better learning of user and item features, as well as for improving recommendation accuracy \cite{mcauley2013hidden}.

\textbf{Enhancing Explainability: }
Explainability is a significant issue in recommender systems. A recommender system needs to be able to provide a rationale for its recommendations to a user. Traditional rating-based systems lack explainability because they solely rely on latent factor learning. Latent factors represent the hidden aspects that influence user preferences, but they usually cannot explain why a user gave a specific rating to an item. Analyzing reviews can identify the specific aspects/features that motivated users to provide certain ratings. Many models have demonstrated that utilizing reviews enhances explainability \cite{mcauley2013hidden,seo2017interpretable}.

%% file: ReviewBasedRec.tex
\section{Review-based recommender systems}

The main components of recommender systems are user/item representation learning, user-item feature interaction mechanism, and overall rating or preference prediction. Traditional recommender systems take user $ID$ and item $ID$ as input and learn their features using latent factor models \cite{koren2009matrix}. However, these static feature-learning methods suffer from data sparsity and scalability problems. It is often beneficial to use as much information as possible, including the user review text\cite{chin2018anr}, images \cite{yilma2023together}, numerical rating scores \cite{cheng20183ncf}, video \cite{mu2023multimodal}, and audio \cite{bogdanov2013semantic}. Many e-commerce platforms such as Amazon, Yelp, and TripAdvisor collect user feedback as textual reviews in addition to numerical ratings. User reviews are a rich source of feedback for learning user preferences and item attributes. Because users are free to express their opinions, it is assumed that their fine-grained preferences are available in the reviews. These reviews can be utilized in various ways to solve problems in traditional rating-based recommender systems, e.g., data sparsity, and explainability problems.

\begin{figure}[ht]
\centering
\includegraphics[width=0.9\textwidth]{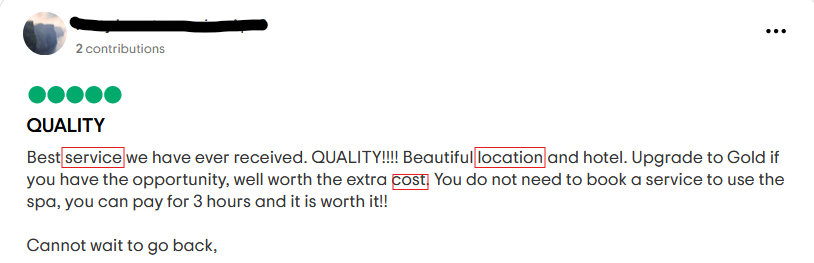}

    \caption{An example of a hotel review highlighting aspects such as \enquote{service}, \enquote{location} and \enquote{cost}}
\label{fig:fig0}
\end{figure}

Various strategies are applied to effectively utilize the reviews in recommendation systems. Based on how reviews are integrated into recommendation systems, we classify review-based recommender systems into four main categories: 1) Generic Review-based Methods, 2) Aspect-based Methods, 3) Review and Rating Fusion Methods, and 4) Ratings and Aspects Fusion Methods. Each category is discussed in details below. Additionally, relevant citations for each category are summarized in Table 1.

\subsection{Generic Review-based Methods}
 The generic review-based method focuses on leveraging user reviews to improve recommendation accuracy. In this case, reviews are used in two different ways, a) User and Item Representation \cite{seo2017interpretable, shuai2022review, almahairi2015learning}, and b) Sentiment Analysis \cite{preethi2017application}. 

\textbf{a) User and Item representations}: 
This approach primarily aims to learn comprehensive representations of both users and items from reviews, without emphasizing any particular aspect or topic. To illustrate this category, suppose a user set $U=\{u_{1}, u_{2}, ..., u_{|U|}\}$, item set $I=\{i_{1},i_{2},...,i_{|I|}\}$, review set $E=\{e_{1},e_{2},...,e_{|E|}\}$, and an overall rating set $R=\{r_{1},r_{2},...,r_{|R|}\}$. In a traditional rating-based recommender system, user $u's$ preference $\hat{r}$ towards item $i$ is estimated by computing the dot product between the user vector and item vector, as shown below:

\begin{eqnarray}
\hat{r}=P_{u}^\mathsf{T} Q_{i}+b
\end{eqnarray}
where $P_{u}$ and $Q_{i}$ are the latent feature representation of user $u$ and item $i$ and $b$ is the bias. $P_{u}$ and $Q_{i}$ are learned from $U$ and $I$ respectively, e.g.,  $P_{u}=f(U)$ and $Q_{i}=f'(I)$. Various algorithms can be used for the implementation of functions $f$ and $f'$, and in some cases, they may use the same algorithm, e.g., DeepCoNN \cite{zheng2017joint}

In a generic review-based system,  a user document $d_{u}$ is created by concatenating all reviews the user has written. Similarly, an item document $d_{i}$ is created by combining all reviews for that item. The features are extracted from these documents using NLP techniques (e.g., GloVe \cite{pennington2014glove}, Word2Vec \cite{church2017word2vec}, or BERT \cite{DBLP:journals/csur/WangHTWHLB24}), and latent representations of users and items are learned. For example, \cite{seo2017interpretable} uses attention mechanisms and convolution layers to learn user and item representations, while \cite{wang2018word} employs a CNN for context-aware review representations.  If the user and item representation learned from user reviews $d_{u}$ and item reviews $d_{i}$ are $X_{u}$ and $X_{i}$ respectively, e.g., $X_{u}=f(d_{U})$ and $X_{i}=f'(d_{I})$, then the equation (3) for rating prediction tasks can be updated as follows:
\begin{eqnarray}
\hat{r}=X_{u}^\mathsf{T} X_{i}+b
\end{eqnarray}
where $b$ is the bias term.

\textbf{b) Sentiment Analysis}: The sentiment analysis approach relies on extracting the user's overall sentiment on the item, and the recommendations made based on sentiment score. Some studies use external sentiment analyzers, while others apply machine learning for sentiment analysis. For example, \cite{asani2021restaurant, preethi2017application, chatterjee2021jumrv1,sumaia2023experimental} use WordNet to extract user sentiment, while \cite{preethi2017application} applies deep learning to capture sentiment directly from reviews.

\subsection{Aspect-based Methods}

Aspect-based methods aim to capture users’ fine-grained opinions by extracting or learning latent representations of specific aspects. Consider an aspect set $A=\{1, 2,3, ..|A| \}$ where $|A|$ is the number of aspects. $A_{U}$ and $A_{I}$ represent user and item aspects respectively. The aspect-based representation of user and item are $X_{u}^{A}$ and $X_{i}^{A}$ respectively. $X_{u}^{A}$ and $X_{i}^{A}$ are a function of $A_{u}$ and $A_{i}$, such that $X_{u}^{A}=f(A_{u})$ and $X_{i}^{A}=f'(A_{i})$. For the aspect-based method, the equation (3) can be reformulated as 

\begin{eqnarray}
\hat{r}=(X_{u}^{A})^\mathsf{T} X_{i}^{A}+b
\end{eqnarray}
where $b$ is the biased term. The term \enquote{aspect} refers to the specific attributes of products. In Figure \ref{fig:fig0} \enquote{Location}, \enquote{service}, and \enquote{cost} represent the aspects. Naturally, a user might not be interested in all aspects of a product but one or more of them. Furthermore, the significance of an aspect to a user can differ from one item to another. Hence, it is vital to capture the aspects and their relevance to users. Aspect-aware recommender models leverage advanced NLP techniques such as topic modeling, deep neural networks, and attention mechanisms to dynamically model user preferences on various aspects.

\subsection{Rating and Review Fusion Methods}
In rating and review fusion methods, both rating-based and review-based features are combined to improve user and item representations. The mathematical representation for this type of method can be derived from equations (4) and (5) as follows.
 \begin{eqnarray}
\hat{r}=\zeta(P_{u},X_{u}) ^\mathsf{T} \zeta'(Q_{i},X_{i})+b
\end{eqnarray}
where $\zeta$ and $\zeta'$ are the aggregation method integrating rating and review based features. They can be same or different for user and item. Rating and review fusion methods also extend to sentiment-focused recommendation systems. In this case, user sentiments are extracted and integrated into the user-item rating matrix to enrich the rating matrix. For example, \cite{osman2021integrating} incorporates the sentiment information into the user-item rating matrix, enhancing the quality of the user-item rating matrix. 
 
\subsection{Ratings and Aspects Fusion Methods}
Ratings and aspects fusion methods integrate rating-based user and item representations with aspect-based representations. The combined effect improves the accuracy of rating predictions or enhances the ranking quality of the recommendation system. Similarly, the equation for this method can be derived from (4) and (5) as follows:
\begin{eqnarray}
\hat{r}=\zeta(P_{u},X_{u}^{A}) ^\mathsf{T} \zeta'(Q_{i},X_{i}^{A})+b
\end{eqnarray}
Different algorithms are used as aggregation functions including, machine learning, probabilistic, counterfactual, etc. For example, \cite{wei2024multi} uses contrastive learning to incorporate graph-based features from ratings to review-based features. Similarly, \cite{wang2023learning} utilizes element-wise product as an aggregation method to integrate rating-based and review-based features.

Some other work focuses on extracting aspect-based sentiment scores and leverages them to collaborative filtering, enhancing recommendation accuracy \cite{hernandez2019comparative,al2020unsupervised,osman2021integrating,al2023trusted,sumaia2023experimental}. On the other hand, \cite{huang2020personalized} computes the aspect-specific sentiment score and searches for similarity-based nearest neighbors for recommendation.

\begin{table*}
\scriptsize
\centering
\caption{Publications of different Categories of review-based papers}
\begin{tabular}{|p{4.0cm}|p{10.0cm}|}
\hline
\textbf{Categories} & \textbf{Publications} \\
\hline
Generic Review-based& \cite{yang2023based,xv2022lightweight,choi2022based,rafailidis2019adversarial,wang2018word,seo2017interpretable,catherine2017transnets,zheng2017joint,kim2016convolutional,almahairi2015learning,liu2023knowledge,wu2021enhanced,jin2020racrec,osman2021integrating,preethi2017application,asani2021restaurant,chatterjee2021jumrv1}\\
\hline
Aspect-based & \cite{li2023prompt,tan2021counterfactual,xiong2021counterfactual,yang2021aspect,liu2021multilingual,ni2019justifying,guan2019attentive,chin2018anr,bauman2017aspect,he2015trirank,li2015learning,shi2022dualgcn,Liu2022ReviewPolarity,Sun2021AnUnsupervised,Ray2020Ensemble,li2020aspect,Huang2020PersonalizedReview,da2020weighted,Da’u2019SentimentAware}\\
\hline
Rating \& Review Fusion & \cite{wei2024multi,chiang2023shilling,shuai2022review,luo2021aware,zhang2021sifn,lyu2021reliable,wang2021leveraging,liu2021learning,gao2020set,liu2020heterogeneous,dong2020asymmetrical,xia2019leveraging,wu2019reviews,liu2019daml,liu2019nrpa,alexandridis2019free,wu2019context,chen2019dynamic,deng2018neural,mei2018attentive,chen2018neural,lu2018coevolutionary,chen2018adversarial,zhang2017joint,xi2021deep,wei2023expgcn,cai2022ri,du2021based,liu2021toward,Liu2021HybridNN,shalom2019generative,zhuang2024improving,xu2018adjective,shoja2019customer,hung2020integrating,hu2020reviewer,osman2021integrating,liaoaspect}\\
\hline

Ratings \& Aspects Fusion & \cite{wang2023multi,wang2023learning,pena2020combining,hyun2018review,cheng2018aspect,cheng20183ncf,tan2016rating,liaoaspect}\\
\hline
\end{tabular}
\label{table:tab1}
\end{table*}

As shown in Table 1, the most popular methods are Rating and Review fusion methods, followed by Aspect-based methods. The least popular method is the rating and aspect fusion method. Overall, integrating review-based features with rating-based methods is found to be used more often.

%% file: Baselines.tex
\section{State-of-the-art Review-based Recommender Systems}

In this section, we briefly present the state-of-the-art review-based recommender systems. Our analysis emphasizes the methodology, the rationale behind each method, features, datasets, evaluation metrics, and performance. We classify these systems into three main categories based on the methods they used: 1) Probabilistic and Topic Modeling Approaches, 2) Deep Learning-Based Approaches, and 3) Miscellaneous. This classification enables a comprehensive understanding of the diverse methods used in review-based recommender systems, ensuring that all major techniques are covered.

\subsection{Probabilistic and Topic Modeling Approaches}
The main focus of the probabilistic approaches is to leverage the topic or aspect features learned from reviews and integrate them with the rating-based features, enhancing the overall recommendation accuracy, scalability, and interpretability. Latent Dirichlet Allocation (LDA) \cite{blei2003latent} and Gaussian Mixture Models (GMM) \cite{reynolds2009gaussian} are the two most successful techniques utilized to achieve this goal.

Most probabilistic approaches, such as the Hidden Factors as Topics (HFT) \cite{mcauley2013hidden} and Rating-Boosted Latent Topics (RBLT)  \cite{tan2016rating}, share a common methodology that integrates the latent dimension in ratings with topics. This hybrid approach is particularly effective in addressing data sparsity by enhancing the system's ability to make accurate predictions even with limited user history. Both models leverage reviews to refine the recommendation process, achieving improved effectiveness across diverse datasets such as Amazon and Yelp.

One of the unique features of HFT \cite{mcauley2013hidden} is its ability to extract the topics from the review that can explain the variation present in ratings and reviews. In addition, this model demonstrates a high prediction accuracy, particularly for new products and users with limited rating history, compared to matrix factorization and LDA-based models. It can also identify the most useful and informative reviews, ensuring interpretability. However, it faces scalability challenges due to its computational demands. In contrast, RBLT \cite{tan2016rating} bridges the advantage of the latent factor model and topic model to integrate the ratings and textual reviews, enhancing the recommendation accuracy. Similar to HFT, RBLT alleviates data sparsity and improves interpretability by exploring item recommendability and user preference distributions in a shared topic space.

The Neural Gaussian Mixture Model (NGMM) \cite{deng2018neural} uses a Gaussian mixture model in its architecture to refine rating predictions. This model utilizes neural networks to represent user and item characteristics through latent vectors, capturing the details of preferences and properties. Its use of a Gaussian Mixture Model at a shared layer enables precise rating predictions, enhancing the integration of textual review analysis into recommender systems for improved accuracy.

The Aspect-Aware Latent Factor Model (ALFM) \cite{cheng2018aspect} correlates latent topics extracted from reviews with latent factors from ratings, focusing on identical aspects. It surpasses traditional models by avoiding direct one-to-one correspondence between topics and factors, which improves both accuracy and interpretability, but it depends heavily on the quality of the review content. The Matching Distribution by Reviews (MDR) model \cite{shalom2019generative} effectively combines deep learning with collaborative filtering for text processing, capturing the distribution of user-item matches found in reviews. It provides better prediction accuracy and reduced runtime across multiple datasets. Finally, the Topic Initialized Latent Factor Model (TIM) \cite{pena2020combining} enhances recommendations accuracy by initializing user and item embeddings within a topic space derived from reviews. This method not only speeds up model convergence but also improves interpretability by aligning embeddings with identifiable topics, offering a more transparent and understandable recommender system.

\begin{table*}[t]
\tiny
\centering
\caption{The Detailed Summary of the Probabilistic Approaches}
\begin{center}
\begin{tabular}{p{1.7cm}p{0.5cm}p{1.5cm}p{1.7cm}p{2.4cm}p{1.7cm}p{3.0cm}}\hline

 Models & Year & Methodology & Features & Datasets & Evaluation & Tasks \\

\hline
\scriptsize HFT \cite{mcauley2013hidden}& \scriptsize 2013 &\scriptsize LDA& \scriptsize ratings \& \scriptsize reviews  &  \scriptsize amazon, \scriptsize yelp \& \scriptsize beer& \scriptsize MSE \& MAE  & \scriptsize accuracy, interpretability \\ 
\scriptsize LMLF \cite{almahairi2015learning}& \scriptsize 2015 &\scriptsize LDA& \scriptsize reviews  &  \scriptsize amazon & \scriptsize MSE  & \scriptsize accuracy \\ 
\scriptsize LBSN \cite{li2015learning}& \scriptsize 2015 &\scriptsize LDA& \scriptsize  reviews  &  \scriptsize yelp \& \scriptsize tripadvisor& \scriptsize MSE  & \scriptsize accuracy\\
\scriptsize RBLT \cite{tan2016rating}  & \scriptsize 2016 &\scriptsize LDA & \scriptsize ratings \& \scriptsize reviews  & \scriptsize amazon  & \scriptsize MSE, P \& R &\scriptsize accuracy\\
\scriptsize ParVecMF \cite{alexandridis2019free}&\scriptsize 2019& \scriptsize PMF &\scriptsize ratings \& reviews &\scriptsize amazon, yelp &\scriptsize MAP, MRR &\scriptsize accuracy \\ 
\scriptsize ALFM \cite{cheng2018aspect} &\scriptsize 2018&\scriptsize LDA &\scriptsize reviews &\scriptsize amazon \& yelp &\scriptsize RMSE &\scriptsize accuracy, interpretability \\ 
\scriptsize NGMM \cite{deng2018neural} &\scriptsize 2018&\scriptsize NN, GMM& \scriptsize reviews &\scriptsize amazon &\scriptsize MSE  & \scriptsize accuracy \\ 
\scriptsize ARAOE \cite{hernandez2019comparative}&\scriptsize 2018&\scriptsize LDA & \scriptsize reviews &\scriptsize amazon, yelp&\scriptsize P, Recall, nDCG &\scriptsize accuracy, time complexity\\
 \scriptsize MDR \cite {shalom2019generative}&\scriptsize2019 &\scriptsize LDA & \scriptsize \scriptsize reviews  & \scriptsize amazon \& yelp  &\scriptsize MSE &\scriptsize accuracy, interpretability\\
 
 \scriptsize TIM \cite {pena2020combining}&\scriptsize2020 &\scriptsize LDA, MF & \scriptsize ratings \& \scriptsize reviews  & \scriptsize amazon \& tripadvisor  &\scriptsize Recall &\scriptsize accuracy, interpretability\\
\hline 

\end{tabular}
\end{center}
\label{tab:probabilistic}
\end{table*}
While all these models demonstrate enhanced recommendation accuracy, their effectiveness varies by application domains. For instance, HFT and RBLT are particularly suitable for new products and users in the e-commerce domain, where user history might be limited. In contrast, NGMM and ALFM show broader applicability by also enhancing recommendations in specialized markets and for unique item groups. The detailed summary of the probabilistic models is presented in Table \ref{tab:probabilistic}.
\subsection{Deep Learning-based Approaches}

Deep learning, a subset of machine learning, involves training large neural networks to recognize patterns in data. The idea of a neural network which is based on neurons came from biological neurons. A simple neural network is made up of multiple layers, and each layer has multiple neurons. The initial and final layers of a network are defined as input and output layers where layers in between are referred to as hidden layers. Neurons in each layer are connected to successive layers and they collect signals from previous layers. The signals on each neuron coming from the previous neurons are accumulated before transmitting to the next layer, and an activation function controls the output of each neuron \cite{Goodfellow-et-al-2016}. The activation function processes the data in a non-linear fashion, making deep learning models capable of learning a complex representation of data. It is a widely used method of learning the representation of text, images, audio, and video signals. An architecture of a simple neural network also called Multi-Layer Perceptrons (MLPs) is shown in Figure \ref{fig:nn}.

\begin{figure}[ht]  
    \centering  
    \begin{tikzpicture}[scale=0.9]

        \foreach \i in {1,2,3} {
            \node[circle, draw, minimum size=1cm] (I-\i) at (0,2.5-\i*1.5) {};
            \node at (-0.7, 2.5-\i*1.5) {$x_{\i}$};
        }
        \node at (0, 2) {Input Layer};

        \foreach \i in {1,2,3,4} {
            \node[circle, draw, minimum size=1cm] (H-\i) at (3,2.75-\i*1.25) {};
            \node at (3.7, 2.75-\i*1.25) {$h_{\i}$};
        }
        \node at (3, 2.5) {Hidden Layer};

        \foreach \i in {1,2} {
            \node[circle, draw, minimum size=1cm] (O-\i) at (6,2.25-\i*1.5) {};
            \node at (6.7, 2.25-\i*1.5) {$y_{\i}$};
        }
        \node at (6, 1.8) {Output Layer};

        \foreach \i in {1,2,3} {
            \foreach \j in {1,2,3,4} {
                \draw[->] (I-\i) -- (H-\j);
            }
        }

        \foreach \i in {1,2,3,4} {
            \foreach \j in {1,2} {
                \draw[->] (H-\i) -- (O-\j);
            }
        }

    \end{tikzpicture}
    \Description{A Simple Multi-Layer Perceptron (MLP) with 3 Layers}
    \caption{A Simple Multi-Layer Perceptron (MLP) with 3 Layers} 
    \label{fig:nn}
\end{figure}

Deep learning architectures such as Convolutional Neural Networks (CNNs), Recurrent Neural Networks (RNNs), Long Short-Term Memory (LSTM) networks, Gated Recurrent Unit (GRU), Attention mechanisms, Graph Neural Networks (GNN), Large Language Models (LLMs) and Autoencoders have revolutionized various fields, including image and speech recognition, natural language processing, and recently, recommender systems. In recommender systems, deep learning-based models have shown an exceptional ability to extract deep, non-linear relationships between users and items, significantly improving recommendation quality. These models can handle vast amounts of data, learning from user interactions, and contextual information to provide personalized recommendations. 
\begin{figure}[ht]
\centering
\includegraphics[width=0.7\textwidth]{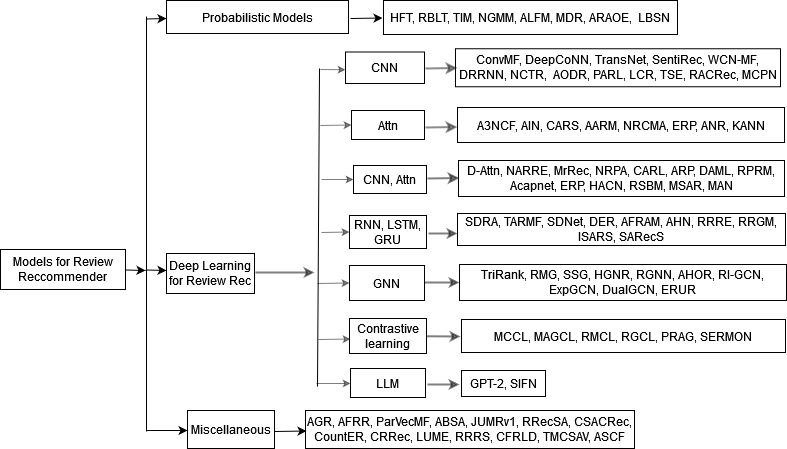}
    \Description{Summary of the Different Review-based Recommender Models.}
    \caption{Summary of the Different Review-based Recommender Models}
\label{fig:fig3}
\end{figure}
By analyzing the rich textual data in reviews, these systems uncover nuanced user preferences and item features that traditional methods might miss.  In this subsection, we briefly summarize state-of-the-art deep learning-based models and analyze the key features along with their effectiveness. Some papers may apply multiple deep learning techniques in their recommender systems, and in this case, we categorize them based on the main models they have used.

\subsubsection{\textbf {Convolutional Neural Network (CNN)-based Approaches:}  }
CNN-based models typically excel in addressing challenges associated with sparse user-to-item interactions by extracting the contextual features from textual reviews. CNNs can identify and capture local features, such as phrases and sentences, that carry significant semantic meaning. This capability allows these models to make more accurate predictions, even with limited data. Most CNN-based approaches, such as \cite{kim2016convolutional,zheng2017joint}, combine traditional Matrix Factorization methods with CNN's power to enhance the overall effectiveness of recommender systems.

The Convolutional Matrix Factorization (ConvMF) \cite{kim2016convolutional} integrates CNNs with probabilistic matrix factorization to effectively capture detailed context within item descriptions, moving beyond the limitation of traditional bag-of-word techniques. ConvMF is particularly noted for its ability to improve recommendation accuracy in conditions of extreme data sparsity. However, its complexity and the use of pre-trained word embeddings indicate potential areas for further refinement and optimization. Utilizing dual neural networks,  the Deep Cooperative Neural Networks (DeepCoNN) \cite{zheng2017joint} learns item properties and user behaviors from review texts. Its shared layer mimics factorization machine techniques, allowing to capture the interactions between latent factors of users and items.   
Expanding on the concept of network representation, the Translation-Based Network Representation Learning (TransNet) \cite{catherine2017transnets} introduces vertex interactions in network representation learning as translation operations. This method allows TransNet to capture richer semantic information, enhancing the social relation extraction task and providing deeper insights into vertex relationships.

A novel sentiment-guided review-aware recommendation model (SentiRec) \cite{hyun2018review} efficiently processes review data by transforming reviews into sentiment-encoded vectors. This significantly reduces computational demands and captures essential sentiment details, enabling tailored recommendations based on emotional content. Additionally, the Word-Driven and Context-Aware Review Modeling for Recommendation (WCN-MF)  model \cite{wang2018word} enhances text analysis by integrating Deep Latent Dirichlet Allocation with CNNs, allowing for detailed extraction of contextual features from reviews, which traditional pooling methods might miss. This method improves the model’s contextual understanding and recommendation relevance.

The Weighted Aspect-based Opinion Mining using Deep Learning for Recommender Systems (AODR) \cite{da2020weighted}
focuses on extracting weighted, aspect-based opinions, effectively identifying the relative importance users assign to different product aspects, thereby aligning recommendations more closely with user preferences. A hybrid Neural network to Combine Textual information and Rating for recommendation (NCTR) \cite{Liu2021HybridNN} integrates CNNs with collaborative filtering to address data sparsity, uncovering latent user-item interactions, and enriching the model's understanding of complex preference patterns. Finally, Deep Rating and Review Neural Network for Item Recommendation (DRRNN) \cite{xi2021deep} leverages both ratings and reviews during training, ensures a comprehensive preservation of semantic information, which enhances personalization.

Together, these models demonstrate the flexibility of CNNs in adapting to different aspects of recommendation challenges, from general feature extraction to coarse-grained and fine-grained sentiment analysis, highlighting their broad applicability across diverse recommendation scenarios.

\subsubsection{\textbf{Neural Attention-based Approaches: }} Attention mechanisms within recommender systems significantly advance traditional algorithms by focusing on the most relevant parts of the data. This method effectively differentiates crucial information from noise, which enhances the interpretability and relevance of the recommendations. Broadly, these models excel in managing the detailed information present in user reviews, which significantly boosts the specificity and personalization of recommendations.

The Adaptive Aspect Attention Model for Rating Prediction (A3NCF) \cite{cheng20183ncf} 
uses an aspect-based attention framework that selectively focuses on various aspects of items as valued by users, extracting preferences and item characteristics from reviews. This offers a refined and aspect-oriented recommendation model. This specificity is beneficial in tailoring recommendations, though A3NCF may face challenges when reviews are sparse or superficial, affecting its ability to fully capture user preferences. Similarly, the Attentive Interaction Network (AIN) \cite{mei2018attentive} extends these capabilities to Context-Aware Recommender Systems (CARS) \cite{kulkarni2020context}  with a multi-module approach that includes interaction-centric, user-centric, and item-centric components. This comprehensive framework allows AIN to dynamically adjust the contextual relationships between users and items, providing deeply personalized recommendations tailored to current user situations. However, AIN might struggle in environments where contextual information is either too generic or excessively varied. 

The Attentive Aspect Modeling for Review-aware Recommendation (AARM)  \cite{guan2019attentive}  specifically tackles the aspect-sparsity issue prevalent in many reviews. By learning from both consistent and varying aspect connections between users and products, AARM significantly enhances the depth of understanding in user-item relationships. Meanwhile, Cross-Modality Mutual Attention (NRCMA), \cite{luo2021aware} utilizes a two-tower structure to synchronize user and item representations through mutual attention, effectively enhancing the match between user preferences and item characteristics, though it faces challenges in very dynamic environments. The review-based recommender with attentive properties (RRAP) \cite{lei2023influence} integrates review properties and sequential information to improve recommendation quality. This model effectively reduces prediction error and enhances interpretability by leveraging attention mechanisms. However, the model relies on historical data and struggles with unseen reviews, which can limit its generalizability. 

Lastly, the Enhanced review-based Rating Prediction (ERP) \cite{wu2021enhanced} incorporates a mechanism to evaluate the general influence of reviewers, allowing it to differentiate and tailor recommendations based on how much trust to place in each review. This capability ensures that the recommendations are not only personalized but are also aligned with the most credible and relevant user feedback, enhancing the accuracy and relevancy.

These attention-based models enhance recommender systems by improving how user data is interpreted and utilized. While all models enhance traditional methods by incorporating advanced attention mechanisms, they vary in applications and challenges. A3NCF focuses on aspect-based ratings, AIN adapts to contextual variations, AARM addresses review aspect sparsity, and NRCMA bridges user-item representation gaps.

\subsubsection{\textbf{Integrated CNN and Attention-based Approaches:}} This approach combines the strengths of CNNs and attention mechanisms to enhance the accuracy of the recommender systems. By utilizing CNNs, this approach effectively extracts and learns deep, complex features from textual data, such as user reviews or item descriptions. The incorporation of attention mechanisms further refines the process by selectively focusing on the most relevant parts of the data, ensuring that the most informative aspects are emphasized. This integrated approach enables a more detailed understanding of user preferences and item characteristics, leading to more accurate and personalized recommendations.  

The Interpretable Convolutional Neural Networks with Dual Local and Global Attention (D-Attn) \cite{seo2017interpretable} utilizes a sophisticated dual attention mechanism that enables a detailed analysis of textual data. By merging local and global attention, D-Attn not only highlights crucial phrases and contextual indicators within reviews but also provides a more comprehensive understanding of the content. This dual-layered attention significantly enhances interpretability over traditional models like Matrix Factorization, HFT \cite{mcauley2013hidden}, and Convolutional Matrix Factorization, making it proficient in managing complex datasets where detailed textual analysis is critical. In contrast,
the Neural Attentional Rating Regression with Review-level Explanations (NARRE) \cite{chen2018neural} focuses on the qualitative value of reviews rather than just numerical ratings. Its unique attention mechanism assesses the usefulness of each review, providing detailed explanations that enhance user trust and deepen satisfaction. NARRE proves to be particularly effective in environments that depend heavily on rich user feedback for recommendation refinement, making it ideal for platforms that prioritize depth and explanation in user interactions.

Context-Aware User-Item Representation Learning (CARL) \cite{wu2019context} and Dual Attention Mutual Learning (DAML) \cite{liu2019daml} both excel in dynamically adapting recommendations based on user and item interactions. CARL leverages CNNs enhanced with an attention layer to derive insights from both review and interaction data, enriching the contextual understanding necessary for precise personalization. DAML integrates local and mutual attention mechanisms, which balance the representation of users and items to ensure accuracy and relevance in its outputs. Neural Recommendation with Personalized Attention (NRPA) \cite{liu2019nrpa} personalizes recommendations by focusing on the informativeness of reviews for different users and items. The Aspect-aware Neural Review Rating (ARP) \cite{wu2019arp} focuses on aspect-level opinions in reviews, providing a detailed understanding of user preferences and item qualities. By using a collaborative learning framework to train both review-level and aspect-level rating predictors, ARP improves rating prediction accuracy, particularly for items with sparse ratings. 

Multilingual Review-aware Deep Recommender System (MrRec) \cite{liu2021multilingual}  tackles the complexities of handling reviews in multiple languages by integrating an aspect-based sentiment analysis module with a multilingual recommendation framework. This model efficiently processes and leverages fine-grained user-item interactions from diverse linguistic backgrounds. To effectively capture the semantic features of reviews, a CNN is used to perform convolution operations on each language and aspect-specific matrix. Additionally, attention mechanisms are utilized to emphasize significant features and interactions. Meanwhile, RPRM \cite{wang2021leveraging} utilizes review properties to assess their usefulness, introducing two new loss functions and a negative sampling strategy that directly link these properties with user preferences. This method allows RPRM to provide a more refined modeling of user preferences and item attributes.  

Overall, the integration of CNNs with attention mechanisms in recommender systems enhances the precision and personalization of recommendations by effectively analyzing complex textual data. 

\subsubsection{\textbf{Recurrent Neural Networks (RNNs)-based Approaches: }} RNN and their variants, such as LSTM, and GRUs, are particularly effective in review-based recommender systems. By leveraging their ability to process sequential information, these models can analyze the detailed sentiment and progression within user reviews. This facilitates a deeper understanding of preferences and item characteristics from textual feedback. Consequently, RNNs enable recommender systems to generate more accurate and personalized recommendations by addressing complex details in user feedback.

The Sentiment-aware Deep Recommender System with neural Attention network (SDRA) \cite{da2019sentiment} further enhances the recommendations by fully leveraging user-generated textual reviews. Integrating a semi-supervised topic model with a deep learning framework (e.g., LSTM) through a neural attention mechanism, it adeptly captures both product aspects and the associated user sentiments from textual reviews, leading to improved recommendation performance. The use of LSTM encoders for capturing contextual word information and a co-attention mechanism for fine-tuning user-item interactions based on these insights further enhances performance. SDRA offers a pathway for incorporating user sentiment and aspect information into the recommendation process, showcasing its potential for more nuanced and effective recommendations. The Topical Attention Regularized Matrix Factorization (TARMF) \cite{lu2018coevolutionary}  integrates attention-based GRUs with matrix factorization, tackling the inherent sparsity in user-item ratings that frequently challenges collaborative filtering systems. TARMF's hybrid approach not only enhances sentiment analysis of product features but also adds a layer of transparency and explainability, setting it apart from traditional matrix factorization methods. The Selective Distillation Network (SDNet) \cite{chen2018adversarial} introduces a distillation approach where the system uses external knowledge, such as user reviews, to train a  \enquote{teacher} model that guides a more streamlined \enquote{student} model that focuses on rating prediction from user-item pair. The \enquote{teacher} model uses both CNN and LSTM for review processing to extract the enriched features from review. This separation allows for efficient runtime operation without sacrificing the depth of knowledge extraction, addressing the dual needs of performance and efficiency in dynamic environments.

Leveraging Ratings and Reviews with Gating Mechanism for Recommendation (RRGM) \cite{xia2019leveraging} and Dynamic Explainable Recommender (DER) \cite{chen2019dynamic}  both emphasize the dynamic nature of user preferences. While the former uses dual attention-based GRU networks to adapt recommendations based on current user-review interactions, DER employs a time-aware GRU to track evolving preferences, supplemented by a CNN for item profiling. These models excel in environments where user preferences are continually changing, offering recommendations that are not only timely but also explainable. Aspect-based Fashion Recommendation with Attention Mechanism (AFRAM) \cite{li2020aspect}, focusing on the fashion industry, processes both local and global aspect features to provide precise fashion advice, demonstrating its capability to enhance recommendation accuracy in a specialized market with unique consumer demands. 

Lastly, Asymmetrical Hierarchical Networks (AHN) \cite{dong2020asymmetrical} and Reliable Recommendation with Review-level Explanations (RRRE) \cite{lyu2021reliable} focus on the granularity and reliability of reviews respectively. AHN addresses the varying scope of user and item reviews by applying differently focused attention modules, enhancing the relevancy of content extracted for recommendations. Specifically, the model utilizes bi-directional LSTM networks to process the sequential data within reviews, allowing for the effective capture and integration of contextual and temporal relationships between sentences. Conversely, RRRE prioritizes the authenticity of reviews, using mechanisms to assess and filter content based on reliability, which is crucial for maintaining the integrity of recommendation outputs.
Overall, these RNN-based models improve recommendation accuracy and user satisfaction by processing sequential review data. They vary in their specific applications—some enhance interpretability and trust (TARMF and RRRE), while others focus on adapting to changing user conditions (DER and RRGM). The use of RNNs alongside technologies like attention and adversarial networks indicates the potential for ongoing advancements in the field, aiming to align recommender systems more closely with user expectations and preferences.

\subsubsection{\textbf{Graph Neural Network (GNN)-based Approaches:}} GNNs  can model the complex user-item interactions in recommendation systems. By exploiting the structural connectivity of data, GNNs excel in capturing high-order relationships. This is particularly effective for addressing data sparsity, cold start problems, and improving the interpretability of recommendations through detailed insights from user and item interactions. Tripartite Graph Ranking (TrinRank) \cite{he2015trirank} evolves beyond traditional collaborative filtering by modeling ternary relationships in a heterogeneous tripartite graph (user-item-aspect). The model focuses on vertex ranking in the tripartite graphs, tailored for personalized recommendation which provides explainability and transparency by modeling aspects in reviews. In contrast, Review Meet Graphs (RMG) \cite{wu2019reviews}  utilizes a multi-view framework that merges review texts with user-item graphs within a GNN architecture. By doing so, RMG captures detailed user-item interactions more adeptly, highlighting the advantage of integrating diverse data views to boost the effectiveness of recommender systems.

Heterogeneous Graph Neural Recommender (HGNR) \cite{liu2020heterogeneous} and Set-Sequence-Graph (SSG) \cite{gao2020set} both address the fundamental challenges of recommender systems. HGNR focuses on alleviating the cold start problem by using social information and reviews to predict links between users and items, demonstrating GNNs' capability to leverage heterogeneous data. Conversely, SSG enhances recommendation accuracy by combining sequential and graphical views, which provides a comprehensive understanding of short-term user preferences and broader collaborative signals.

Review Graph Neural Network (RGNN) \cite{liu2020heterogeneous}  tackles bias and noise in data by constructing individual review graphs for each user/item pair. This model emphasizes the importance of understanding both local and global word dependencies in reviews, thus enhancing the precision of rating predictions through a sophisticated graph attention mechanism and personalized pooling strategies. Finally, Aspect-Aware Higher-Order Representations (AHOR) \cite{wang2023learning} utilize heterogeneous graphs to capture aspect-aware relationships, enriching the learning of high-order user and item representations through multi-hop connectivity. The incorporation of an attention mechanism to selectively focus on relevant features ensures that AHOR not only enhances performance but also significantly improves the interpretability of recommendations.

Overall, these GNN-based models utilize graph structures to analyze relationships but vary in their approach—some focus on broad interaction patterns, while others concentrate on detailed aspects of content and context. This variety highlights the potential for developing hybrid models that could merge these approaches to enhance accuracy and customization in recommendations.
\begin{figure}[ht]
\centering
\includegraphics[width=0.6\textwidth]{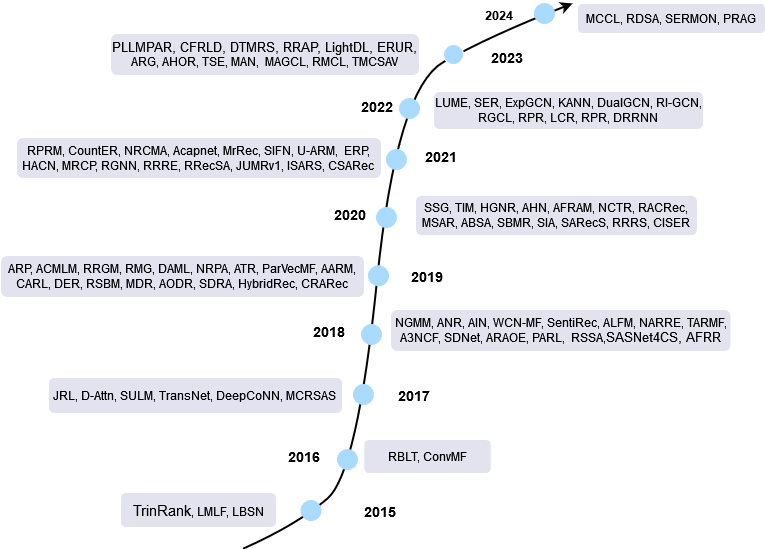}
    \Description{The Evolution of the Review-based Recommendation Between 2015 to 2024}
    \caption{The Evolution of the Review-based Recommendation Between 2015 to 2024}
\label{fig:evolve}
\end{figure}

\subsubsection{\textbf{Contrastive Learning-based Approaches:}} The Contrastive Learning-based Approach \cite{khosla2020supervised,xia2022hypergraph} refines recommender systems by its unique power of data representation learning technique - modeling the data representation by minimizing the distance between similar data points while maximizing the distance between dissimilar data points. This approach has been successful in tasks requiring detailed data understanding, and making recommendations based on user reviews. The review-aware Graph Contrastive Learning (RGCL) framework leads this approach by enhancing user and item embeddings through the integration of review data. It utilizes contrastive learning techniques to improve node embedding and interaction modeling, thus addressing data sparsity and more accurately capturing user preferences. Similarly, the Review-based Multi-intention Contrastive Learning (RMCL) \cite{yang2023based} method introduces learning intention representations from reviews based on a mixed Gaussian distribution hypothesis. It uses a multi-intention contrastive learning strategy to link user and item reviews, fostering a deeper connection between them.

The Multi-Aspect Graph Contrastive Learning (MAGCL) \cite{wang2023multi}  advances the application of contrastive learning in GNNs by incorporating multi-aspect graph representations and multi-task learning. This model breaks down user-generated reviews into distinct semantic aspects, enhancing the granularity of user preference analysis. Its contrastive learning module not only improves the model’s predictive accuracy but also helps to overcome the common over-smoothing problem in GNNs. Recently, Multi-Level Cross-modal Contrastive Learning (MCCL) \cite{wei2024multi} explores contrastive learning across different modalities, enhancing the ability of models to leverage semantic information in a self-supervised way.

Overall, these models utilize contrastive learning to refine how recommender systems process review data using diverse and specific methods and different complexities. RGCL and RMCL focus on enhancing node and intention understanding, respectively, while MAGCL and MCCL expand the scope to multi-aspect and cross-modal analysis.

\subsubsection{\textbf{LLMs-based Approaches:}} Large Language Models (LLMs) excel at processing complex text, which enhances the personalization and clarity of recommender systems \cite{laskar2023systematic,laskar2023can}. A significant advancement in this area has been achieved by integrating LLMs, specifically GPT-2, into the recommendation process \cite{li2023prompt}. By utilizing a prompt tuning technique, the model dynamically extracts key aspects reflecting user preferences and item properties directly from textual reviews. These extracted aspects are then used to inform and personalize the recommendation process, leading to a more tailored and insightful user experience. 

This integration not only improves the relevance and quality of recommendations but also contributes to the transparency and explainability of the system, addressing a critical need in recommender systems research. Evaluated on three large-scale datasets, the model demonstrates superior performance over existing methods, showcasing its potential for practical application and scalability. This study demonstrates the transformative potential of LLMs in recommender systems, paving the way for further exploration and development in this exciting intersection of natural language processing and personalized recommendation.
The Sentiment-aware Interactive Fusion Network (SIFN) \cite{zhang2021sifn} addresses the precision of sentiment interpretation in recommendations by utilizing Bidirectional Encoder Representation From Transformer (BERT) \cite{kenton2019bert} for encoding and a lightweight sentiment learner for semantic analysis. This model enhances the accuracy of item suggestions by using explicit sentiment labels, although it faces challenges with sparse or vague reviews.

{
\tiny 
\begin{longtable}{p{1.7cm}p{0.4cm}p{1.9cm}p{1.7cm}p{2.7cm}p{2.3cm}p{3.0cm}} 
\caption{The Detailed Summary of the Deep Learning-based Approaches\label{tab:deeplearning}}\\
\toprule
 Models & Year & Architecture & Features & Datasets & Evaluation & Tasks \\
\midrule
\endfirsthead

\multicolumn{7}{c}%
{{\bfseries \tablename\ \thetable{} -- continued from previous page}} \\
\toprule
 Models & Year & Methodology & Features & Datasets & Evaluation & Tasks \\
\midrule
\endhead

\midrule \multicolumn{7}{r}{{Continued on next page}} \\ \midrule
\endfoot

\bottomrule
\endlastfoot

 \scriptsize TriRanK \cite{he2015trirank}&\scriptsize 2015 &\scriptsize GNN  &\scriptsize reviews &\scriptsize amazon &\scriptsize HR, NDCG   &\scriptsize transparency, explainability\\ 

 \scriptsize ConvMF \cite{kim2016convolutional}&\scriptsize 2016 &\scriptsize CNN, MF  &\scriptsize reviews &\scriptsize amazon, MovieLens &\scriptsize RMSE   &\scriptsize accuracy\\ 

 \scriptsize DeepCoNN \cite{zheng2017joint}&\scriptsize 2017&\scriptsize CNN &\scriptsize reviews &\scriptsize yelp &\scriptsize MSE   &\scriptsize  accuracy\\ 
\scriptsize TransNet \cite{catherine2017transnets}&\scriptsize 2017&\scriptsize CNN &\scriptsize reviews &\scriptsize yelp &\scriptsize MSE  &\scriptsize accuracy \\ 
 \scriptsize JRL \cite{chen2018adversarial} &\scriptsize 2017&\scriptsize MLP &\scriptsize rat, img \& rev &\scriptsize amazon &\scriptsize NDCG &\scriptsize accuracy \\ 
 \scriptsize RDSA \cite{preethi2017application} &\scriptsize 2017 &\scriptsize RNN &\scriptsize reviews &\scriptsize amazon &\scriptsize F1 &\scriptsize accuracy\\ 

\scriptsize D-Attn \cite{seo2017interpretable}&\scriptsize 2017 &\scriptsize CNN, Attn  &\scriptsize reviews &\scriptsize yelp \& amazon &\scriptsize  MSE & \scriptsize accuracy\\ 

\scriptsize SULM \cite{bauman2017aspect}&\scriptsize 2017 &\scriptsize \text{SULM}  &\scriptsize reviews  &\scriptsize yelp &\scriptsize  Precision@K & \scriptsize accuracy, interpretability\\

\scriptsize NARRE \cite{chen2018neural} &\scriptsize 2018 &\scriptsize CNN, Attn  &\scriptsize reviews &\scriptsize yelp \& amazon &\scriptsize  MSE &\scriptsize accuracy, explainability\\ 
 
 \scriptsize SentiRec \cite{hyun2018review} &\scriptsize 2018&\scriptsize CNN &\scriptsize reviews &\scriptsize amazon &\scriptsize MSE &\scriptsize scalability, memory efficient \\
 
 \scriptsize WCN-MF \cite{wang2018word} &\scriptsize 2018 &\scriptsize CNN, LDA, MF &\scriptsize reviews &\scriptsize amazon &\scriptsize RMSE &\scriptsize accuracy\\

\scriptsize A3NCF \cite{cheng20183ncf}&\scriptsize 2018&\scriptsize Attn&\scriptsize ratings \& reviews &\scriptsize amazon \& yelp &\scriptsize MSE &\scriptsize accuracy\\

\scriptsize TARMF \cite{lu2018coevolutionary}&\scriptsize 2018& \scriptsize GRU, Attn &\scriptsize ratings \& reviews & \scriptsize amazon \& yelp &\scriptsize MSE & \scriptsize transparency, explainability\\
\scriptsize AIN \cite{mei2018attentive}&\scriptsize 2018&\scriptsize Attn&\scriptsize  ratings \& reviews &\scriptsize yelp, food \& frappe &\scriptsize RMSE &\scriptsize accuracy, interpretability\\
\scriptsize ANR \cite{chin2018anr}&\scriptsize 2018&\scriptsize CNN & \scriptsize reviews &\scriptsize yelp \& amazon&\scriptsize MSE, MAE &\scriptsize accuracy, interpretability\\

\scriptsize PARL \cite{wu2018parl}&\scriptsize 2018&\scriptsize CNN & \scriptsize reviews &\scriptsize beer \& amazon&\scriptsize MSE &\scriptsize accuracy\\

\scriptsize MCPN \cite{tay2018multi}&\scriptsize 2018&\scriptsize Attn & \scriptsize reviews &\scriptsize yelp \& amazon&\scriptsize MSE, MAE &\scriptsize accuracy, interpretability\\

 \scriptsize SDNet \cite{chen2018adversarial} &\scriptsize 2018&\scriptsize CNN, LSTM &\scriptsize reviews&\scriptsize amazon &\scriptsize MSE &\scriptsize accuracy, time complexity \\ 
\scriptsize CARL \cite{wu2019context}&\scriptsize 2019&\scriptsize CNN, Attn & \scriptsize reviews &\scriptsize amazon &\scriptsize MSE &\scriptsize accuracy, interpretability\\
\scriptsize NRPA \cite{liu2019nrpa}&\scriptsize 2019& \scriptsize CNN&\scriptsize reviews &\scriptsize amazon &\scriptsize MSE &\scriptsize accuracy\\
\scriptsize DAML \cite{liu2019daml}&\scriptsize 2019&\scriptsize CNN, MF&\scriptsize reviews &\scriptsize amazon &\scriptsize MAE &\scriptsize accuracy, interpretability\\
\scriptsize AARM \cite{guan2019attentive}&\scriptsize 2019&\scriptsize Attn &\scriptsize  reviews &\scriptsize yelp \& amazon&\scriptsize N \& HR &\scriptsize accuracy, interpretability\\
 \scriptsize RSBM \cite{Huang2020PersonalizedReview} &\scriptsize 2019 &\scriptsize CNN, Attn &\scriptsize reviews &\scriptsize amazon \& yelp &\scriptsize MSE &\scriptsize accuracy, scalability\\
 \scriptsize AODR \cite{da2020weighted} &\scriptsize 2019 &\scriptsize CNN &\scriptsize reviews &\scriptsize amazon &\scriptsize MSE, RMSE &\scriptsize accuracy\\
\scriptsize ARP \cite{wu2019arp}&\scriptsize 2019 &\scriptsize Attn & \scriptsize ratings \& reviews &\scriptsize chines res. reviews&\scriptsize MAE &\scriptsize accuracy\\

 \scriptsize SDRA \cite{da2019sentiment} &\scriptsize 2019 &\scriptsize LSTM &\scriptsize reviews &\scriptsize amazon \& yelp &\scriptsize MSE &\scriptsize accuracy\\

 \scriptsize ATR \cite{rafailidis2019adversarial}&\scriptsize 2019& \scriptsize Adversarial Net &\scriptsize reviews &\scriptsize amazon &\scriptsize MSE &\scriptsize accuracy \\ 

 \scriptsize ACMLM \cite{ni2019justifying}&\scriptsize 2019& \scriptsize GNN &\scriptsize reviews &\scriptsize amazon, yelp &\scriptsize RMSE &\scriptsize accuracy \\ 

\scriptsize RRGM \cite{xia2019leveraging}&\scriptsize 2019& \scriptsize GRU, Attn &\scriptsize reviews &\scriptsize amazon, yelp &\scriptsize MSE &\scriptsize accuracy, interpretability\\ 

 \scriptsize RMG \cite{wu2019reviews}&\scriptsize 2019& \scriptsize GNN &\scriptsize ratings \& reviews &\scriptsize amazon, yelp &\scriptsize RMSE &\scriptsize accuracy \\ 

\scriptsize DER \cite{chen2019dynamic}&\scriptsize 2019& \scriptsize CNN, GRU &\scriptsize ratings \& reviews &\scriptsize amazon, yelp &\scriptsize RMSE &\scriptsize accuracy, explainability \\ 

\scriptsize CRARec \cite{shoja2019customer}&\scriptsize 2019& \scriptsize LDA, MLP &\scriptsize ratings \& reviews &\scriptsize amazon &\scriptsize MSE &\scriptsize accuracy \\ 

 \scriptsize RACRec \cite{jin2020racrec} &\scriptsize 2020 &\scriptsize CNN &\scriptsize reviews &\scriptsize amazon &\scriptsize RMSE &\scriptsize accuracy\\

 \scriptsize AFRAM \cite{li2020aspect} &\scriptsize 2020 &\scriptsize CNN, LSTM &\scriptsize reviews &\scriptsize amazon &\scriptsize MSE, MAE &\scriptsize accuracy\\ 
 \scriptsize HGNR \cite{liu2020heterogeneous} &\scriptsize 2020 &\scriptsize GNN &\scriptsize reviews &\scriptsize yelp, L., eps &\scriptsize MSE, MAE &\scriptsize accuracy\\ 
 \scriptsize SSG \cite{gao2020set} &\scriptsize 2020&\scriptsize CNN &\scriptsize reviews&\scriptsize amazon \& yelp &\scriptsize HR, NDCG &\scriptsize accuracy \\ 
 \scriptsize SARecS \cite{hung2020integrating} &\scriptsize 2020&\scriptsize CNN-LSTM &\scriptsize ratings \& reviews &\scriptsize amazon &\scriptsize RMSE &\scriptsize accuracy \\ 

  \scriptsize AHN \cite{dong2020asymmetrical} &\scriptsize 2020&\scriptsize Bi-LSTM, Attn &\scriptsize ratings \& reviews &\scriptsize amazon &\scriptsize MAE &\scriptsize accuracy, interpretability\\ 

 \scriptsize EHRecSA \cite{Ray2020Ensemble} &\scriptsize 2020&\scriptsize BERT &\scriptsize reviews&\scriptsize amazon \& yelp &\scriptsize P, R, F &\scriptsize accuracy \\ 

  \scriptsize MSAR \cite{peng2020mutual} &\scriptsize 2020&\scriptsize CNN, Attn &\scriptsize ratings \& reviews  &\scriptsize amazon &\scriptsize MSE &\scriptsize accuracy \\ 
 \scriptsize NCTR \cite{Liu2021HybridNN} &\scriptsize 2020 &\scriptsize CNN &\scriptsize ratings \& reviews &\scriptsize amazon \& movie &\scriptsize MSE, MAE &\scriptsize accuracy\\ 
\scriptsize RPRM \cite{wang2021leveraging}&\scriptsize 2021&\scriptsize CNN&\scriptsize reviews &\scriptsize amazon &\scriptsize P, R \& M &\scriptsize accuracy\\
\scriptsize NRCMA \cite{luo2021aware}&\scriptsize 2021 &\scriptsize CNN& \scriptsize reviews &\scriptsize amazon &\scriptsize MSE &\scriptsize accuracy\\
\scriptsize Acapnet \cite{yang2021aspect} &\scriptsize 2021&\scriptsize CNN &\scriptsize reviews &\scriptsize amazon &\scriptsize MSE &\scriptsize accuracy, interpretability\\ 

\scriptsize ISARS \cite{dang2021approach} &\scriptsize 2021&\scriptsize CNN, LSTM &\scriptsize ratings \& reviews &\scriptsize amazon &\scriptsize RMSE, MAE, NMAE &\scriptsize accuracy \\ 

\scriptsize MRCP \cite{liu2021toward} &\scriptsize 2021&\scriptsize CNN, Attn &\scriptsize ratings \& reviews &\scriptsize amazon, yelp &\scriptsize MSE &\scriptsize accuracy \\ 

 \scriptsize MrRec \cite{liu2021multilingual}&\scriptsize 2021 &\scriptsize CNN, Attn &\scriptsize reviews &\scriptsize amazon \& goodreads&\scriptsize MSE &\scriptsize accuracy, interpretability\\
\scriptsize SIFN \cite{zhang2021sifn} &\scriptsize 2021&\scriptsize Attn &\scriptsize reviews&\scriptsize amazon &\scriptsize MSE &\scriptsize accuracy \\ 

\scriptsize U-ARM \cite{sun2021unsupervised} &\scriptsize 2021 &\scriptsize MLP &\scriptsize reviews &\scriptsize amazon \& yelp &\scriptsize RMSE &\scriptsize accuracy, explainability \\ 

 \scriptsize ERP \cite{wu2021enhanced} &\scriptsize 2021 &\scriptsize CNN &\scriptsize ratings \& reviews &\scriptsize amazon \& yelp &\scriptsize MSE, RMSE &\scriptsize accuracy \\ 

\scriptsize HACN \cite{wu2021enhanced} &\scriptsize 2021 &\scriptsize CNN, Attn &\scriptsize reviews &\scriptsize amazon &\scriptsize MAE, RMSE &\scriptsize accuracy\\ 

\scriptsize RGCL \cite{shuai2022review} &\scriptsize 2022 &\scriptsize GNN, CL &\scriptsize reviews &\scriptsize amazon \& yelp &\scriptsize MSE &\scriptsize accuracy\\ 

\scriptsize SER \cite{choi2022based} &\scriptsize 2022 &\scriptsize MLP &\scriptsize reviews &\scriptsize amazon \& yelp &\scriptsize MSE, NDCG &\scriptsize accuracy\\

 \scriptsize DRRNN \cite{xi2021deep} &\scriptsize 2022 &\scriptsize CNN, MLP &\scriptsize ratings \& reviews&\scriptsize amazon &\scriptsize RMSE &\scriptsize accuracy \\ 

  \scriptsize KANN \cite{liu2023knowledge} &\scriptsize 2022 &\scriptsize Attn &\scriptsize reviews &\scriptsize amazon, Imdb &\scriptsize RMSE &\scriptsize accuracy, explainability\\ 

  \scriptsize RI-GCN \cite{cai2022ri} &\scriptsize 2022 &\scriptsize Graph, CNN &\scriptsize ratings \& reviews &\scriptsize amazon \& yelp &\scriptsize MSE &\scriptsize accuracy \\ 

  \scriptsize DualGCN \cite{shi2022dualgcn} &\scriptsize 2022 &\scriptsize Attn &\scriptsize reviews &\scriptsize amazon \& yelp &\scriptsize N, MSE, P, R &\scriptsize accuracy \\ 

 \scriptsize RPR \cite{Liu2022ReviewPolarity} &\scriptsize 2022 &\scriptsize CNN &\scriptsize reviews &\scriptsize amazon \& yelp &\scriptsize MSE \& MAE &\scriptsize accuracy, explainability\\ 

  \scriptsize MAN \cite{yang2023man} &\scriptsize 2022 &\scriptsize CNN, Attn &\scriptsize reviews &\scriptsize amazon &\scriptsize RMSE \& MAE &\scriptsize accuracy\\ 

 \scriptsize AHOR \cite{wang2023learning} &\scriptsize 2023 &\scriptsize GNN &\scriptsize reviews &\scriptsize amazon \& yelp &\scriptsize NDCG, HR &\scriptsize accuracy\\ 
 \scriptsize ExpGCN \cite{wei2023expgcn} &\scriptsize 2023 &\scriptsize Graph, CNN &\scriptsize ratings \& reviews &\scriptsize amazon, yelp, trip, Imdb &\scriptsize P, R, N&\scriptsize accuracy, explainability\\

 \scriptsize MAGCL \cite{wang2023multi} &\scriptsize 2023 &\scriptsize Graph \& CL &\scriptsize  ratings \& reviews &\scriptsize amazon \& yelp &\scriptsize MRR, NDCG, HR &\scriptsize accuracy, interpretability\\ 

  \scriptsize ARG \cite{chiang2023shilling} &\scriptsize 2023 &\scriptsize RL &\scriptsize ratings \& reviews &\scriptsize amazon \& yelp &\scriptsize Rel, GF, AE  &\scriptsize accuracy\\ 

  \scriptsize LightDL \cite{chiranjeevi2023lightweight} &\scriptsize 2023 &\scriptsize RL &\scriptsize reviews &\scriptsize amazon \& yelp &\scriptsize P, R, F  &\scriptsize accuracy\\ 

  \scriptsize RMCL \cite{yang2023based} &\scriptsize 2023 &\scriptsize CL &\scriptsize reviews &\scriptsize amazon  &\scriptsize MSE  &\scriptsize accuracy\\ 
 
 \scriptsize TSE \cite{yang2023two} &\scriptsize 2023 &\scriptsize CNN &\scriptsize reviews &\scriptsize amazon &\scriptsize MSE, RMSE, MAE &\scriptsize accuracy\\ 
 
  \scriptsize ERUR \cite{liu2023enhancing} &\scriptsize 2023 &\scriptsize Graph &\scriptsize reviews &\scriptsize amazon &\scriptsize MSE, MAE &\scriptsize accuracy\\ 
  \scriptsize RRAP \cite{lei2023influence} &\scriptsize 2023 &\scriptsize CNN, LSTM, BERT &\scriptsize reviews &\scriptsize amazon \& yelp &\scriptsize MSE &\scriptsize accuracy, interpretability\\ 

\scriptsize PLLMPAR \cite{li2023prompt} &\scriptsize 2023 &\scriptsize LLM &\scriptsize reviews &\scriptsize amazon \& yelp &\scriptsize MSE, MAE &\scriptsize accuracy\\ 

\scriptsize DTMRS \cite{gheewala2024exploiting} &\scriptsize 2024 &\scriptsize MLP &\scriptsize reviews &\scriptsize amazon, LibraryThing, yelp &\scriptsize MSE, RMSE, MAE &\scriptsize accuracy\\ 

\scriptsize MCCL \cite{wei2024multi} &\scriptsize 2024 &\scriptsize GNN \& CL &\scriptsize reviews &\scriptsize amazon \& yelp &\scriptsize NDCG, P &\scriptsize accuracy, interpretability\\ 

\scriptsize SERMON \cite{liaoaspect} &\scriptsize 2024 &\scriptsize CL &\scriptsize ratings \& reviews &\scriptsize amazon, yelp \& tripadvisor &\scriptsize MSE, BL, RG, BS &\scriptsize accuracy, explainability\\ 

\scriptsize PRAG \cite{zhuang2024improving} &\scriptsize 2024 &\scriptsize CL &\scriptsize ratings \& reviews &\scriptsize amazon, yelp \& tripadvisor &\scriptsize RMSE &\scriptsize accuracy, explainability\\ 

\end{longtable}
}
\vspace{-5.67mm} 
{\footnotesize Note: L = LibraryThings, Eps = Epinions, Rel = Relevance, AE = Attack Effectiveness, BL = BLEU Score, RG = ROUGE, BS = BERTScore}

\begin{table*}[htbp!]
\tiny 
\centering
\caption{The Detailed Summary of the Miscellaneous}
\begin{center}
\begin{tabular}{p{1.7cm}p{0.5cm}p{2.5cm}p{1.7cm}p{2.2cm}p{2.0cm}p{3.0cm}}\hline

 Models & Year & Methodology & Features & Datasets & Evaluation & Tasks \\
\hline

\scriptsize AFRR \cite{xu2018adjective}&\scriptsize 2018&\scriptsize NNF & \scriptsize ratings \& reviews &\scriptsize Imdb \& amazon &\scriptsize MAP  &\scriptsize sparsity \& transparency\\

\scriptsize contextCF \cite{osman2021integrating} &\scriptsize 2021&\scriptsize Similarity &\scriptsize ratings \& reviews &\scriptsize amazon &\scriptsize MAE, RMSE, R &\scriptsize sparsity \& sensitibity \\ 

\scriptsize CountER \cite{tan2021counterfactual} &\scriptsize 2021 &\scriptsize Counterfactual &\scriptsize ratings \& reviews &\scriptsize amazon \& yelp &\scriptsize NDCG, HR, F1 &\scriptsize accuracy \& explanation\\ 

\scriptsize CRRec \cite{xiong2021counterfactual} &\scriptsize 2021 &\scriptsize Counterfactual &\scriptsize reviews &\scriptsize amazon \& yelp &\scriptsize NDCG, HR, F1 &\scriptsize accuracy \\ 
\scriptsize LUME \cite{xv2022lightweight} &\scriptsize 2022 &\scriptsize Knowledge Dis &\scriptsize reviews &\scriptsize amazon \& yelp &\scriptsize MSE, NDCG &\scriptsize accuracy \& biases\\ 
 \scriptsize RRRS \cite{preethi2017application} &\scriptsize 2020&\scriptsize Similarity &\scriptsize review&\scriptsize amazon \& yelp &\scriptsize MAE &\scriptsize accuracy \& reliability \\

\scriptsize TMCSAV \cite{al2023trusted} &\scriptsize 2023 &\scriptsize Similarity &\scriptsize reviews &\scriptsize amazon &\scriptsize MSE, RMSE, MAE &\scriptsize accuracy \& trust\\ 

\scriptsize ASCF \cite{al2023cluster} &\scriptsize 2023 &\scriptsize K-means clustering &\scriptsize reviews &\scriptsize amazon \& yelp &\scriptsize MSE, RMSE, MAE &\scriptsize accuracy\\ 

\hline 

\end{tabular}
\end{center}
\label{tab:Miscellaneous}
\end{table*}

\vspace{-5.67mm} 
{\footnotesize Note: NNF = Nearest Neighbour Frequency, Dis = Distillation, Cosine Similarity measure}

\subsection{\textbf{Miscellaneous Approaches}} 
This section explores a range of methodologies beyond traditional probabilistic and deep learning approaches. These methods use unique strategies to leverage user reviews, addressing aspects not fully covered in the previously discussed categories. Examples include semantic similarity search, sentiment extraction, counterfactual arguments, reinforcement learning, and nearest neighbor search.

Reinforcement Learning, an agent learns to make decisions by interacting with an environment \cite{wiering2012reinforcement}. The agent receives rewards or penalties based on its actions, which guide it toward optimizing its behavior over time. In review-based recommender systems (RBRS), RL can be particularly effective in scenarios where user preferences are dynamic and the system must adapt to continuously evolving data. RL algorithms can optimize the recommendation process by learning from user interactions and feedback, such as reviews, to enhance the personalization and accuracy of the recommendations \cite{afsar2022reinforcement}.

A recent paper explores the use of RL for generating shilling attacks in RBRS, highlighting a novel vulnerability where fake reviews are used to manipulate recommendation outcomes. The Shilling Black-box Review-based Recommender Systems \cite{chiang2023shilling}, utilize adversarial training to test and enhance system robustness against fake reviews that manipulate recommendation outcomes. This approach not only demonstrates the system's susceptibility to shilling attacks but also its capacity to defend against such tactics, emphasizing the need for robust security measures in recommender systems.

Counterfactual Explainable Recommendation (CountER) \cite{tan2021counterfactual} leverages counterfactual reasoning to provide clear and meaningful explanations for recommendations by identifying minimal changes in item aspects that could influence user decisions. CountER not only enhances the transparency of recommendations but also assesses the impact and clarity of its explanations, significantly outperforming other methods in explainability. 

The Lightweight Unbiased Multi-teacher Ensemble (LUME) \cite{xv2022lightweight} addresses computational complexity, biased recommendations, and poor generalization by implementing a multi-teacher ensemble with debiased knowledge distillation. This strategy aggregates insights from various pretrained models to form a streamlined, unbiased system that excels in generalization. 

On the other hand, some other works focus on utilizing external tools for sentiment score extraction from the reviews, and leveraging them to collaborative filtering method for rating prediction \cite{musto2017multi,osman2021integrating,al2020unsupervised,huang2020personalized}. For instance, a multi-criteria recommendation system \cite{musto2017multi} utilizes user reviews to extract aspect-specific sentiment scores through opinion mining, integrating them into collaborative filtering algorithms to enhance prediction accuracy. The Sentiment Aspect-Based Retrieval Engine (SABRE) \cite{caputo2017sabre} uses a two-level granularity approach, identifying aspects and sub-aspects within text segments. It performs aspect-based sentiment classification and enables filtering during the exploration of retrieved documents, facilitating fine-grained opinion retrieval. ContextCF \cite{osman2021integrating} uses four lexicon-based word lists to extract the context-aware sentiments score from the review, combining them to rating based collaborative filtering for improving the recommendation quality. Additionally, both the Average Trusted Multi-Criteria Similarity (TMCSAV) and Aspect Clustering for Collaborative Filtering (ACCF) \cite{al2020unsupervised}  model employ SEAE \cite{al2021various} for semantically enhanced aspect extraction, and generates associated sentiment scores using a lexicon approach, specifically SentiWordNet. While TMCSAV demonstrates that the integration of aspect specific sentiment score improves the performance of the multicriteria recommendation system across all metrics (e.g. MSE, MAE, and RMSE) ACCF is effective in case of unbalanced large-scale reiviews. Similarly, Aspect Sentiment Similarity-based
 Personalized review Recommendation (A2SPR) \cite{huang2020personalized} generates aspect-specific sentiment score by using SentiWordNet3.0.

In educational settings, a sentiment-based model \cite{hazar2022learner} combines explicit and implicit ratings from user comments to address data sparsity and improve recommendation accuracy. A trusted user model \cite{al2023trusted}, enhances multi-criteria recommendations by incorporating adjective features, reducing prediction errors, and building trust with users. An experimental study \cite{sumaia2023experimental} further combines implicit sentiment and aspect information with explicit ratings, demonstrating improved CF performance on large-scale datasets. Additionally, a clustering approach groups similar aspects extracted from reviews to reduce data sparsity, enhancing CF performance in movie recommendations \cite{al2023cluster}. Sentiment analysis also facilitates a distributed recommendation system \cite{singh2024sentiment} that processes large-scale data for more precise recommendations. In agriculture, a recommendation model \cite{sharma2024sentiment} applies sentiment analysis and clustering to offer disease-specific product recommendations to farmers, illustrating the versatility of these sentiment-based approaches across different sectors.

The Sentiment Utility Logistic Model (SULM) \cite{bauman2017aspect} utilizes sentiment analysis to identify key aspects of products that influence user preferences, enhancing the personalization of recommendations. For example, it might recommend a restaurant specifically for its standout seafood, effectively tailoring suggestions to user preferences. Demonstrated in various settings like restaurants and hotels, SULM’s aspect-focused recommendations have been well-received for improving user satisfaction. Building on this focus on user insights, the Adversarial Training for Review-based Recommendations (ATR) \cite{rafailidis2019adversarial} model integrates user reviews with collaborative filtering. By using adversarial training, this model aligns deep representations of reviews with user and item features, aiming to improve the accuracy of recommendations. The detailed description of this categories in shown in Table \ref{tab:Miscellaneous}.

\textbf{Context-Specific Suitability of Different Approaches: } Each review-based recommender system approach has its own strengths and limitations, making it suitable for specific situations depending on the context and requirements. Probabilistic and topic modeling methods \cite{blei2003latent,reynolds2009gaussian,tan2016rating} are highly interpretable and effective in handling sparse data. This makes them ideal for scenarios like early-stage products or specialized markets where data is limited but interpretability is crucial. However, they may struggle with capturing complex non-linear relationships in extensive datasets. Deep learning approaches, including CNNs, RNNs, and attention mechanisms \cite{kim2016convolutional,zheng2017joint,cheng20183ncf,wu2021enhanced}, perform exceptionally well at modeling intricate patterns and non-linear relationships, making them suitable for platforms with abundant data like large e-commerce sites or streaming services. These methods are particularly effective when high accuracy is required, but they often demand significant computational resources and may lack interpretability. Graph Neural Networks (GNNs) \cite{he2015trirank,wu2019reviews,liu2020heterogeneous}  are well-suited for applications involving rich relational data, such as social networks, where understanding the interactions between users and items is critical. They can capture high-order connectivity but may be complex to implement. Contrastive learning methods \cite{khosla2020supervised,xia2022hypergraph,yang2023based} are advantageous in situations with abundant unlabeled data and are effective for cold-start problems, but they require careful tuning and large datasets to perform well. LLM-based approaches \cite{li2023prompt} are powerful for understanding complex language data and capturing detailed semantic relationships in reviews. They are suitable for applications that require advanced natural language understanding but can be resource-intensive and less interpretable. Therefore, the choice of method should align with the application's specific requirements, such as data availability, the necessity for interpretability, computational constraints, and the complexity of user-item interactions. Table 5 provides a clear summary of the trade-offs between these approaches, highlighting their scalability, interpretability, strengths, limitations, and specific use cases.\\

\begin{table*}
\centering
\caption{Comparison of Different Approaches of Review-based Recommender Systems}
\tiny
\begin{tabular}{|>{\raggedright\arraybackslash}p{2cm}|
                >{\raggedright\arraybackslash}p{1.3cm}|
                >{\raggedright\arraybackslash}p{1.3cm}|
                >{\raggedright\arraybackslash}p{3cm}|
                >{\raggedright\arraybackslash}p{3cm}|
                >{\raggedright\arraybackslash}p{3cm}|}
\hline
\textbf{Approaches} & \textbf{Scalability} & \textbf{Interp-} \newline \textbf{retability} & \textbf{Strengths} & \textbf{Limitations} & \textbf{Specific Use Cases} \\
\hline
\textbf{Probabilistic \& Topic Modeling \cite{mcauley2013hidden,tan2016rating,deng2018neural}} & Medium & High & Effective with sparse data, provides interpretable topics & Limited in capturing non-linear relationships; computationally intensive for large datasets & E-commerce (e.g., new products), where user history is limited \\
\hline
\textbf{CNN-Based \cite{kim2016convolutional,zheng2017joint,xi2021deep}} & Medium & Low-Medium & Extracts local features, captures contextual phrases for improved prediction with sparse data & Requires substantial data and computational resources, interpretability depends on additional techniques & E-commerce sites, especially for products with text-heavy reviews \\
\hline
\textbf{Attention-Based \cite{cheng20183ncf, mei2018attentive,luo2021aware,lei2023influence}} & Medium & High & Selectively emphasizes relevant data for high personalization, improves relevance and interpretability & May struggle with sparse or superficial reviews, effectiveness relies on review quality & Contextual recommendation settings; high-user-interaction platforms \\
\hline
\textbf{Integrated CNN + Attention \cite{seo2017interpretable, chen2018neural,wang2021leveraging}} & Low-Medium & Medium-High & Combines deep feature extraction with selective focus for better recommendation relevance & Complex architectures may increase runtime, challenging to maintain transparency & Personalized recommendations with rich review content \\
\hline
\textbf{RNN-Based (LSTM, GRU) \cite{da2019sentiment, li2020aspect}} & Low & Medium & Processes sequential information for sentiment and aspect progression in reviews & Requires detailed sequential data; challenging to interpret, can be computationally demanding & Sequential analysis settings; sentiment-based product recommendations \\
\hline
\textbf{Graph Neural Networks (GNNs) \cite{he2015trirank,wu2019reviews,liu2020heterogeneous}} & Medium-High & Medium-High & Captures high-order user-item relationships, handles sparsity and complex interactions & Complex data structures; interpretability can be limited without additional techniques & Social and item networks; sparse user interactions \\
\hline
\textbf{Contrastive Learning-Based \cite{khosla2020supervised,xia2022hypergraph}} & Medium & Medium & Models intricate user-item relationships; effective for multi-aspect connections & Limited scalability in large datasets; complex training process & Fine-grained preference modeling; cross-modal applications \\
\hline
\textbf{LLMs-Based  \cite{li2023prompt}} & Low-Medium & Low-Medium & Extracts specific aspects from textual reviews for enhanced personalization and clarity & Data and resource-intensive; interpretability tied to model transparency & Rich review platforms; personalized recommendations \\
\hline
\textbf{Reinforcement Learning-Based \cite{wiering2012reinforcement,chiang2023shilling}} & Low-Medium & Low & Adapts to dynamic user preferences through continuous feedback, useful in changing environments & Requires extensive user interaction data, may struggle with real-time adaptation & Dynamic recommendation systems; user-driven content platforms \\
\hline
\end{tabular}
\label{tab:comparison}
\end{table*}

%% file: Datasets.tex
\section{Datasets and Performance of different Models}

\subsection{Datasets Overview}
Review-based recommender models are evaluated mainly with the datasets that contain user, item, and users' textual review feedback. The detailed statistics and the sources are presented below:

\begin{itemize}
    \item \textbf{Amazon Review\footnote{\url{https://developer.imdb.com/non-commercial-datasets/}}:} Amazon is one of the largest and most commonly used product review datasets, originating from Amazon website. With 21 distinct categories of items, this dataset is the most extensive publicly accessible product review dataset to date, encompassing user ID, item ID, rating, and textual feedback. This dataset includes a variety of niches, listed in the left column of Table 6, ranging from cell phones to electronics.
    
    \item \textbf{Yelp Review\footnote{\url{https://www.yelp.com/dataset}}:} It is the largest restaurant review dataset containing more than 1M reviews and ratings. The dataset is collected from Yelp.com, an online platform for restaurant reviews. This dataset also comes with user, and item IDs along with textual reviews.

    \item \textbf{Beer Review\footnote{\url{https://cseweb.ucsd.edu/~jmcauley/datasets.html\#multi\_aspect}}: } The Beer review dataset is also often used to evaluate the recommendation model. It is collected from the ratebeer.com website. Data was extracted for a period of 10 years, up to November 2011, comprising approximately 3 million reviews

    \item \textbf{TripAdvisor Review\footnote{\url{https://www.cs.virginia.edu/~hw5x/Data/LARA/TripAdvisor/}}: } 
    Tripadvisor.com is a tourism management platform, which provides a range of services such as hotel bookings, flight reservations, restaurant recommendations, cruise options, and car rental facilities. Within this dataset, one can find comprehensive evaluations including overall ratings, multi-criteria assessments, and textual reviews. There exist a total of 8 distinct criteria, each accompanied by its respective rating: value, rooms, location, cleanliness, check-in/front desk, service, sleep quality, and business service. Ratings for both individual criteria and overall experiences fall within a scale from -1 to 5.
    \item \textbf{IMDB\footnote{\url{https://developer.imdb.com/non-commercial-datasets/}}: } It is the world's most popular and authoritative source for movie, TV and celebrity content. The dataset comes with ratings and reviews.
    \item \textbf{Librarythings\footnote{\url{https://cseweb.ucsd.edu/~jmcauley/datasets.html\#social\_data}}: }  LibraryThing is a free, high-quality catalog designed to help users track their reading progress or manage their entire personal library. This dataset also comes with ratings and reviews.
    \item\textbf{Goodreads\footnote{\url{https://www.goodreads.com/genres/dataset}}: } These datasets consist of reviews from the Goodreads book review platform, along with various attributes describing the items. Notably, they capture multiple layers of user interaction, including actions such as adding books to a "shelf," providing ratings, and marking them as read.
    \item\textbf{Epinions\footnote{\url{https://cseweb.ucsd.edu/~jmcauley/datasets.html\#social\_data}}: } This is a general consumer review dataset.

\end{itemize}

\begin{table}
\centering
\caption{Data statistics}
\tiny
\begin{tabular}{|c| c| c| c| c|}
\hline
\textbf{Datasets} & \textbf{Number of users} & \textbf{Number of items} &\textbf{Number of reviews} & \textbf{Density}\\

\hline
Cell Phones & 27879 &10429 &194439 &0.067\%\\
\hline
Digital Music & 5,541 &3,568 & 64,706&0.330\%\\
\hline
Home and Kitchen & 66519 &28237 & 551682&0.03\%\\
\hline
Foods & 14,681 &8,713 & 151,254&0.118\%\\
\hline
Automotive & 2,928 &1,835 & 20,473&0.99\%\\
\hline
Baby &19445 &7050 &160792 &0.122\%\\
\hline
Office Product & 4,905 &2,420 &53,258 &0.449\%\\
\hline
Toys and games& 19,412&11,924 &167,597 &0.072\%\\
\hline
Instant Video& 5,130 &1,685 &37,126 &0.433\%\\
\hline
Musical Instruments &1,429  &900 &10,261 &0.798\%\\
\hline
Video games & 24,303&10,672 &231,577 &0.089\% \\
\hline
Beauty &22,363  &12,101 &198,502 &0.0734\%\\
\hline
Clothing, Shoes, and Jewelry & 39,387 & 23,033&278,677 &0.0307\%\\
\hline
Movies and TV &123,960  &50,052 &1,697,533 &0.0274\%\\
\hline
Sports Outdoors & 35,598 &18,357 &296,337 &0.045\%\\
\hline
Office Products &4,905 &2,420 &53,228&0.449\%\\
\hline
Pet Supplies &18,070 &8,508 &155,692&0.99\%\\
\hline
CDs and Vinyl &75,258 &64,443 & 1,097,494&0.023\% \\
\hline
Tools and Home Improvement & 15,438 &10,217 & 133,414 & 0.99\%\\
\hline
Grocery \& Gourmet Food & 127487 &41320 & 1143470 &0.22\% \\
\hline
Electronics & 2,832 &19,816 &53,295 &0.095\%\\

\hline
TripAdvisor Hotel &429,928 & 3,828& 620,172&0.9996\%\\
\hline
Yelp & 45,981 &11,537 &229,907 &0.0433\%\\
\hline
Beer &40,213 &110,419 &2,924,127 &0.05\%\\
\hline
IMDB &  2088 & 4668 &126,874 &0.98\%\\
\hline
Goodreads &808,749& 1,561,465 &225,394,930 &0.017\%\\
\hline
Librarythings & 53,246 & 167,286 & 1,334,648 &0.015\%\\
\hline
Epinions & 27,545 & 30,324 & 91,739 & 0.003\%\\
\hline
\end{tabular}
\label{tab:dataset}
\end{table}

\subsection{Performance Comparison Across Models and Datasets}
This subsection explores the performance of various models evaluated on common benchmark datasets, highlighting their strengths and limitations. 

The Review-based Multi-Intention Contrastive Learning (RMCL) model has shown superior performance across multiple datasets, as measured by Mean Squared Error (MSE) \cite{yang2023based}. On datasets such as Musical Instrument, Office, Food, Games, and Sports, RMCL outperformed several state-of-the-art models, including NeuMF, ConvMF, DeepCoNN, D-Attn, NARRE, CARL, DAML, MSAR, and RGCL. Notably, RMCL achieved its best performance in the Musical Instruments domain, demonstrating its ability to effectively capture contextual features specific to this dataset. However, its performance in the Food domain was comparatively lower, suggesting that the architectural design of RMCL might struggle with domain-specific details.

Similarly, the Multi-aspect Graph Contrastive Learning (MAGCL) framework exhibited exceptional performance across datasets like Digital Music, Toys and Games, CDs and Vinyl, and Yelp \cite{wang2023multi}. It consistently outperformed other models, such as GC-MC, NGCF, LightGCN, SGL, ALFM, $CF^{2}$, DeepCoNN, NAREE, HGNR, RGCL, and DGCLR. The highest performance was observed on the Yelp dataset, while the Toys and Games dataset reported the least improvement, indicating potential limitations in handling certain types of user-item interactions. These results underline the efficacy of graph contrastive learning-based architectures in achieving both low error rates and superior ranking measures across diverse datasets.

%% file: EvaluationMetrics.tex
\section{Evaluation Metrics}

Research on review-based recommender systems focuses on two core tasks: rating prediction and top-N items recommendation \cite{shalom2021natural}. For the rating prediction task, the problem is formulated as a regression problem and the models are evaluated with MSE, RMSE, and MAE \cite{yang2023based,tay2018multi,da2020weighted,al2023trusted}. In contrast, top-N items recommendation tasks are formulated as a ranking problem, evaluating them with NDCG@N, recall@N, precision@N, F1@N, HR, MRR, and MAP \cite{wei2024multi,wang2023multi,wei2023expgcn,zhang2017joint}. While the former helps to evaluate the user's sentiment prediction accuracy, the latter explores the relevancy of the list of recommended items. Each of the evaluation metrics is defined below.

\subsection{Rating Prediction}
For a given user $u$ and an item $i$, the rating prediction task involves estimating the overall sentiment or overall ratings ($\hat r$), indicating the user $u's$ overall preferences towards the item $i$. If the true rating is $r$, the Mean Squared Error (MSE), Root Mean Squared Error (RMSE), and Mean Absolute Error (MAE) are defined as follows.

MSE measures the squared difference between the prediction and the ground truth, giving more weight to larger errors \cite{shani2011evaluating}. Mathematically, it is defined as follows:
\begin{eqnarray}
MSE=\frac{1}{N} \sum_{i=1}^{N} (\hat{r_{i}} -r_{i})^2 
\end{eqnarray}

Similar to MSE, RMSE is the square root of the average of the squared difference between predictions and the true ratings but disproportionately penalizes the larger errors \cite{chai2014root}.

\begin{eqnarray}
RMSE=\sqrt{\frac{\sum_{i=1}^{N} (\hat{r_{i}} -r_{i})^2}{N} }
\end{eqnarray}

On the other hand, MAE measures the average absolute difference between the prediction and the ground truth. MAE treats all errors equally without emphasizing larger errors more than the smaller ones \cite{willmott2005advantages}.
\begin{eqnarray}
MAE=\frac{1}{N} \sum_{i=1}^{N} |\hat{r_{i}} -r_{i}|
\end{eqnarray}
where $N$ is the total number of data points for evaluations. For imbalanced test sets (where certain items are overrepresented compared to others), it is preferable to measure the MAE or RMSE separately for each item and then take the average because for an imbalanced distribution of items, MAE and RMSE are heavily influenced by the error on a very few frequent items \cite{shani2011evaluating}. In terms of interpretability, MAE is superior to MSE and RMSE. RMSE takes care of interpretability with a stronger penalty for larger errors, while MSE is the least interpretable due to its squared units.

\subsection{Top-N Items Recommendation}
The objective of the top-N recommendation task is to recommend top-N most relevant items to users. In this subsection, we discuss various evaluation metrics for recommendation ranking quality.

\textbf{NDCG (Normalized Discounted Cumulative Gain): }
NDCG is a ranking quality metric, which measures the relevancy and the ranking quality of the recommended items \cite{liu2014towards}.  For top-N items, NDCG is calculated as: 
\setlength{\abovedisplayskip}{1pt} 
\begin{eqnarray}
NDCG@N=\frac{DCG@N}{iDCG@N}
\end{eqnarray}

where $DCG@N$ and $iDCG@N$ measures the Discounted Cumulative Gain and ideal Discounted Cumulative Gain respectively. $\textit{CG (Cumulative Gain)}=\sum_{i=1}^{N} relevanceScore_{i}$, is defined as the sum of the relevance score for all the recommended items, and $\textit{DCG (Discounted Cumulative
Gain)}=\sum_{i=1}^{N} \frac{relevanceScore_{i}}{\log_{2}(i+1)}$, takes into account the positions of the relevant items. It is a widely used evaluation metric that considers both the items' relevancy and their positions in the ranked list \cite{isinkaye2015recommendation}. When both the relevancy and the ranking order matter, NDCG is useful.

\textbf{MRR (Mean Reciprocal Rank): }
MRR focuses on retrieving the first relevant item from the top-N items.
\setlength{\abovedisplayskip}{1pt} 
\begin{eqnarray}
MRR=\frac{1}{U} \sum_{u=1}^{U}\frac{1}{rank_{u}}
\end{eqnarray}
where $U$ is the total number of users and $rank_{u}$ is the rank of the first relevant item for user \textit{u}. It ensures the position of the first relevant item in the ranked list. MRR@N is useful for evaluating how quickly the first relevant result appears. If no relevant item is found within the top N, the reciprocal rank for that user is 0. A higher MRR score indicates better performance in placing relevant items early in the ranking.

\textbf{HR (Hit Ratio): }
Hit Ratio is one of the most widely used top-N evaluation metrics. HR computes the share of users that receive at least one relevant item from the recommended top-N items. HR is computed as: 
\setlength{\abovedisplayskip}{1pt} 
\begin{eqnarray}
HR=\frac{U_{hit}^{L}}{U}
\end{eqnarray}
where $U$ is the total number of users and $U_{hit}^{L}$ is the number of users with at least one relevant recommendation. Hit Ratio ensures that in the top results, there is at least one relevant item in the top results. It is particularly useful when it matters to have any relevant item in the top N recommended items.

\textbf{MAP (Mean Average Precision): } MAP is another popular top-N metric, which measures the average precision across all relevant ranks within the top-N recommended items in the list. It is computed as follows,
\setlength{\abovedisplayskip}{1pt} 
\begin{eqnarray}
MAP=\frac{1}{|U|}\sum_{u=1}^{|U|} AP(u)
\end{eqnarray}
where AP(u) is the average precision for the user $u$ in the ranked list. Average Precision (AP) at position N can be defined as: $AP@N=\frac{1}{m}\sum_{k=1}^{N} precision (k) \times rel(k) $ where $m$ is the total number of relevant items for a particular user. Precision(k) is the precision on position k and rel(k) is 1 if the item at position k is relevant otherwise it is 0.

\textbf{Precision@N: } Precision measures how many of the recommended items are relevant. Higher precision is considered better when it is important to have more relevant items in the top-N list \cite{karypis2001evaluation}.

\textbf{Recall@N: } It is the ratio between the number of relevant items in the top N items and the total number of relevant items. When it is important to retrieve as many relevant items as possible within a cutoff, recall is used \cite{karypis2001evaluation}.

\textbf{F1@N: } F1 measures the harmonic mean of precision and recall at top N positions. It is ideal when both precision and recall are equally important.

Furthermore, explainability of recommendation systems is evaluated with BLEU, ROUGE, and BERT scores \cite{dong2017learning,10.1145/3580488}. \textbf{BLEU Score} measures the quality of candidate generation by comparing the overlap of n-grams between the true review and the reference generation \cite{post2018call}. A greater overlap indicates a higher quality. Uni-grams are used to assess the accuracy of individual word generation, while higher-order n-grams help measure the fluency of the sentence. BLEU-N (N=1,4) primarily relies on these n-gram comparisons. \textbf{ROUGE Score} focuses on recall rather than precision \cite{lin2004rouge}. It examines how many of the n-grams from the reference generation appear in the output. On the other hand \textbf{BERT Score} measures the similarity between two contextual embeddings \cite{zhang2019bertscore}. Apart from the above traditional evaluation metrics, the shilling attack is measured using relevance score, generation fluency, and attack effectiveness \cite{chiang2023shilling}.

%% file: Challenge.tex
\section{MAJOR CHALLENGES IN REVIEW-BASED RECOMMENDER SYSTEMS} 

Review-based recommender systems uniquely incorporate user-generated content to predict preferences and suggest items. While these systems leverage detailed insights from textual reviews, they encounter specific challenges. This section outlines these unique challenges, discussing both historical solutions and areas where challenges persist, emphasizing why these issues continue to impact the effectiveness of review-based recommender systems.

\subsection{Representation Learning}
A primary challenge in developing effective review-based recommender systems lies in learning meaningful representations of user behavior and item attributes from reviews. User-generated text is often unstructured and inherently noisy, sometimes containing slang and emojis \cite{chen2015recommender}. Extracting meaningful information from such complex data is challenging \cite {liu2023enhancing,almahairi2015learning, baral2016maps,cheng20183ncf}. Various deep learning and NLP techniques have been applied to model the complexity of unstructured text data \cite{aziz2024corenlp,kumar2024aspect,al2020unsupervised}. Traditional word embedding methods (e.g. Word2Vec \cite{church2017word2vec}, GloVe \cite{pennington2014glove}) are commonly used for word vector representation \cite{aziz2024corenlp,al2020unsupervised,kumar2024aspect}. However, these fixed-size word embedding methods often fail to learn contextual representations for larger context windows and for out of vocabulary words, leading to sub-optimal recommendation performance \cite{seo2017interpretable,zheng2017joint}. 

To this end, Transformer-based models were introduced to capture contextual sensitivity, learning rich representations from longer review sequences \cite{ray2021ensemble}. Another approach is contrastive learning \cite{liu2021self,khosla2020supervised} that focuses on learning user-item representations by pulling positive sample pairs closer and pushing negative pairs apart. These techniques show great potential; however, they still encounter scalability challenges when integrated with larger models \cite{wei2024multi}.

\subsection{Incorporating Reviews into Recommender Systems}
The key to enhancing the performance of review-based recommender systems lies in effectively integrating review features into the recommendation process \cite{chen2015recommender}. Reviews can be incorporated in four main ways: 1) creating user/item profiles from reviews to enrich representations \cite{seo2017interpretable,zheng2017joint}, 2) learning aspect-based user and item features \cite{Liu2022ReviewPolarity}, 3) combining review features with ratings \cite{liu2019daml,liu2019nrpa}, and 4) fusing aspect-based features with rating-based latent features \cite{wang2023multi,wang2023learning}. 

Simply aggregating reviews to construct user or item profiles can overlook the dynamic nature of user preferences, which are influenced by contextual factors like time, location, mood, and trends. Traditional word embedding methods ( e.g., GloVe \cite{pennington2014glove} or Word2Vec \cite{church2017word2vec} ), generate fixed-size embedding for a word and may not capture these details, leading to potential inaccuracies in profiling. For instance, reviews may express different sentiments over time or across various contexts, which means static user profiles may not reflect true preferences. Research such as that in \cite{cheng20183ncf} suggests that using RNN or LSTM models can capture temporal changes in user sentiment and preference, helping address this challenge.

Moreover, when merging review features with ratings, there is a risk of mismatch—sentiments expressed in reviews might differ from the ratings given. For example, a user may give a high rating but write a negative review, creating inconsistencies in the data. To mitigate this, models like Dual Learning \cite{li2020aspect}, which optimize for both sentiment in reviews and numerical ratings, have been proposed to extract latent aspect features about users and items separately through two parallel paths. However, such dual-objective models still need refinement to consistently align textual and numerical data. Another challenge lies in the inherent noise and bias within user reviews, as users can provide contradictory feedback across different products, or even within the same review, creating difficulties in capturing consistent preferences.

Addressing these challenges requires more sophisticated models that account for the temporal and contextual dynamics of user decisions, thereby enhancing both the accuracy and interpretability of the recommendation. A potential solution is to use attention mechanisms or hybrid neural networks that can selectively focus on relevant parts of the reviews while minimizing the influence of contradictory or irrelevant information. This allows the system to distill meaningful insights from reviews despite inconsistencies, improving overall quality.

\subsection{Scalability}
Handling the vast volumes of review data on e-commerce platforms introduces significant scalability challenges. These include managing sparse and sometimes unclear reviews, extracting meaningful insights efficiently, and accommodating the sheer volume of data processed. The scalability of recommender systems is further complicated by the need to adapt to the dynamic nature of user preferences and item popularity, which may shift due to factors like seasonal demand, marketing efforts, or new product launches. Systems must be capable of updating in real-time to reflect these changes, ensuring recommendations remain relevant and effective \cite{xiong2021counterfactual}.

One possible solution to scalability is the use of distributed systems such as Apache Spark, which enable parallel processing of large review datasets. By breaking down the dataset into smaller chunks, these systems can process reviews efficiently in real-time without compromising performance. Furthermore, new research into Graph is promising in terms of scalability. GNNs can efficiently process relationships between users, items, and reviews by leveraging graph structures that handle large-scale data more effectively.

The integration of diverse data types further amplifies scalability challenges \cite{choi2022based}. Modern e-commerce platforms do not rely solely on textual reviews; they also incorporate images, videos, and audio feedback, creating a multi-modal dataset. Each data type requires specific processing techniques to extract meaningful features that can inform the recommendation process. For instance, NLP techniques are suited for textual data, while image processing algorithms are necessary for analyzing user-uploaded photos. Successfully merging these varied data sources demands sophisticated data fusion techniques and machine learning models that can handle multi-modality without compromising the efficiency or scalability of the recommender system.

\subsection{Privacy and Ethical Considerations} 

As review-based recommender systems increasingly rely on user-generated content to enhance personalization and accuracy, addressing privacy, ethical handling, and bias becomes crucial. User reviews often contain personal identifiers and data that, if mismanaged or exposed, could result in privacy breaches. Additionally, text in reviews can naturally carry biases from the user, introducing potential discrimination into the system. Tackling these biases, detecting them, and mitigating their influence within recommendation algorithms are vital challenges. Ethical considerations around data collection, consent, and usage are also paramount to maintain user trust and regulatory compliance. Developing systems that prioritize user privacy through secure data handling, anonymization techniques, and transparent user consent processes is essential. New research in differential privacy is proving to be a viable solution for protecting user data in review-based systems. By adding noise to the data, these models can prevent user identification while maintaining high recommendation quality. Moreover, creating algorithms that prevent discrimination and ensure fairness in recommendations is crucial to foster an inclusive and trustworthy digital environment.

\section{Real-World Applications of Review-Based Recommender Systems}

Review-based recommender systems have proven crucial across various industries, providing enhanced, personalized recommendations by analyzing user-generated reviews \cite{raza2024comprehensive}. Below, we discuss applications in key sectors such as e-commerce, hospitality, media streaming, healthcare, finance, and education.

In \textbf{e-commerce} (e.g., Amazon, Alibaba), these systems provide personalized recommendations by analyzing user sentiments on various product aspects from user-generated reviews \cite{dang2021approach, schafer2007collaborative,hwangbo2018recommendation,chen2015recommender, Huang2022ECommerce,shoja2019customer}. By extracting these insights from user reviews, these systems present highly relevant product suggestions, reducing choice overload for consumers and directly contributing to increased customer satisfaction and conversion rates \cite{adomavicius2010multi}. For instance, Amazon's integration of review-based recommendations has been significantly enhancing users' shopping experiences, allowing them to make more informed purchase decisions based on peer opinions \cite{hwangbo2018recommendation,shoja2019customer}.

In \textbf{media streaming}, services such as Netflix and Hulu use review-based recommender systems to refine content suggestions by analyzing viewer feedback on elements including plot, acting, and visuals \cite{gomez2015netflix,amatriain2013big}. By incorporating sentiment analysis, these platforms are able to recommend shows and movies that better align with individual user tastes, thus boosting user engagement and satisfaction. For example, Netflix’s success in personalizing content through user feedback has been linked to its substantial increase in viewer retention, highlighting the value of incorporating review data in maintaining user loyalty \cite{amatriain2013big}.

In \textbf{healthcare}, review-based recommender systems guide patients in finding healthcare providers that suit their specific needs, using reviews that highlight aspects such as care quality, communication, and professionalism \cite{tran2021recommender,raza2023improving}. Zocdoc uses these systems to guide patients in navigating their options, allowing them to choose personalized providers \cite{siddiqui2013cancellations}. Such systems have demonstrated a positive impact on patient satisfaction by empowering them to make informed choices regarding their healthcare \cite{tran2021recommender}.

In the \textbf{finance sector}, review-based recommender systems enable banks and Fintech platforms to recommend financial products such as credit cards, loans, and savings accounts based on user reviews about fees, interest rates, and customer service \cite{zibriczky122016recommender}. By tailoring financial product suggestions according to customer feedback, they can build stronger customer relationships. For example, certain banks have reported increased customer loyalty by using review insights to match customers with products that better meet their financial needs \cite{musto2015personalized}. 

In \textbf{hospitality}, platforms like TripAdvisor use review-based recommender systems to suggest hotels, restaurants, and travel experiences based on aspects such as service quality, location, and amenities \cite{majid2013context}. This approach enhances the travel planning experience, as users can select accommodations and services that align closely with their preferences. TripAdvisor's use of this type of recommender systems has been noted for increasing user trust and satisfaction, as travelers rely on peer insights to make more informed travel choices \cite{majid2013context}.

Other sectors, such as \textbf{education} and \textbf{travel}, also leverage review-based recommender systems \cite{tan2008learning,kumar2024aspect}. In education, platforms such as Coursera and Udemy recommend courses based on user feedback about teaching quality and course difficulty, helping learners make optimal choices in their educational journeys. Similarly, in travel, apps use reviews to suggest destinations and activities that match user preferences, enhancing their overall experience and satisfaction with tailored recommendations \cite{dang2021approach}.

%% file: Discussion.tex
\section{DISCUSSION AND FUTURE DIRECTIONS}

Despite challenges, review-based recommender systems advance the recommendation process by using fine-grained insights from user reviews to deliver personalized suggestions. These systems explore the textual content of reviews to grasp users' sentiments, preferences, and fine-grained opinions about products or services. This deeper analysis facilitates enhanced personalization, as the systems can comprehend individual tastes and requirements with greater precision, resulting in recommendations that closely align with each user's specific preferences. Moreover, review-based systems improve accuracy by considering qualitative information present in reviews, such as detailed explanations, experiences, and comparisons, which may not be captured by ratings.

Additionally, review-based recommender systems excel at capturing detailed information about products or services, providing recommendations that consider specific attributes or characteristics. This adaptability to diverse preferences ensures that the recommendations can meet various user interests, accommodating different priorities such as quality, price, or specific features. By leveraging user-generated content, review-based systems offer a trustworthy recommendation process, fostering confidence in users' decision-making. 

\subsection{Enhancing Interpretability and Trust}
Future advancements should focus on interpretability and trust. Users benefit not only from accurate recommendations but also from understanding why specific items are recommended.  Enhancing interpretability involves developing models that can explicitly link recommendations to specific aspects mentioned in reviews, such as quality, usability, or aesthetics. This can be achieved through more sophisticated aspect-based analysis and by incorporating explanation modules that can convey the rationale behind recommendations in an easily understandable manner. Building systems that offer transparent reasoning behind their suggestions could significantly improve user trust and satisfaction.

\subsection{Leveraging Multi-modal Data} 
As user reviews increasingly include multimedia content in reviews, such as images, videos, and audio clips, future recommendation systems could benefit from a multi-modal approach that integrates these diverse types of data. This involves not only analyzing textual content but also understanding and extracting sentiments and preferences from images and videos. For instance, visual sentiment analysis from product images or user-generated content in reviews could offer additional insights into user preferences, enhancing the recommendation process. Research in multi-modal learning and its integration into recommender systems remains an under-exploited area with considerable potential for innovation.

\subsection{Leveraging Multi-criteria Rating Data} 
Incorporating multi-criteria ratings, which reflect specific aspects of user satisfaction, can enhance both the performance and interpretability of review-based recommender systems. Tripadvisor\footnote{\url{https://www.cs.virginia.edu/~hw5x/Data/LARA/TripAdvisor/}}, for instance, collects ratings on different criteria including location, Service, Cleanliness, Value, Sleep Quality,  etc. The ratings on the individual criteria represent users' fine-grained preferences because not all the criteria are equally important to a user. Therefore, integrating these criteria ratings into the review and overall ratings could potentially improve the performance and provide better interpretability. Furthermore, multi-criteria rating information is valuable in mitigating the issue of data sparsity. In cases where reviews are sparse, leveraging multiple criteria can aid in learning user preferences. Recently, MRRRec \cite{hasan2022multi}, and 
 \cite{hasan2023aspect} demonstrate the effectiveness of integrating multi-criteria rating data with reviews to enhance the performance of the recommender system. However, a sophisticated approach is required to take full advantage of the utilizing criteria ratings into reviews.
 
\subsection{Addressing Data Sparsity }
Despite the richness of review data, the issues of data sparsity \cite{grvcar2006data} and the cold start problem persist, where new users or items have insufficient interactions to generate accurate recommendations. Future research could explore novel ways to tackle these problems, possibly by leveraging user demographics, and item metadata, or by using transfer learning techniques where knowledge from one domain is applied to another. Additionally, unsupervised and semi-supervised learning \cite{learning2006semi} methods that can effectively utilize sparse data to find patterns could play a vital role in mitigating these challenges. 

\begin{table}[htbp!]
\centering
\caption{Future Directions in Review-based Recommender Systems}
\tiny 
\label{tab:future_directions}
\begin{tabular}{|p{4cm}|p{5.4cm}|p{5.4cm}|}
\hline
\textbf{Area of Focus} & \textbf{Proposed Directions} & \textbf{Impact} \\
\hline
Enhancing Interpretability and Trust & Develop models linking recommendations to aspects in reviews; incorporate explanation modules. & Improves user trust and satisfaction by making recommendation logic transparent and understandable. \\
\hline
Leveraging Multimodal Data & Integrate text with images, videos, and audio to enrich user profiles and item features. & Enhances depth of insights into user preferences and improves recommendation accuracy. \\
\hline
Leveraging Multi-criteria Rating Data & Combine fine-grained aspect ratings with reviews to refine recommendations. & Enhances performance and interpretability, addressing data sparsity issues. \\
\hline
Addressing Data Sparsity and Cold Start & Use user demographics, item metadata, and transfer learning to improve new user/item recommendations. & Mitigates cold start problems and enhances recommendation accuracy for new users/items. \\
\hline
Ethical Considerations and Bias Mitigation & Develop mechanisms to detect and mitigate biases; ensure fairness and diversity. & Builds trust and compliance, ensuring recommendations are equitable and unbiased. \\
\hline
Real-time Adaptation and Personalization & Implement dynamic models that adjust recommendations based on real-time data. & Keeps recommendations relevant and responsive to current user behavior and trends. \\
\hline
Cross-Domain Recommendations & Apply domain adaptation techniques to leverage knowledge across product areas. & Improves recommendation flexibility and consistency across various product categories. \\
\hline
Large Language Models and Recommendations & Use LLMs for analyzing reviews and generating relevant suggestions. & Leverages advanced NLP capabilities to enhance recommendation quality and relevance. \\
\hline
Responsible Recommendations & Focus on privacy, diversity, and fairness in system design and operation. & Ensures that recommendation systems operate responsibly, promoting user trust and societal benefit. \\
\hline
\end{tabular}
\end{table}

\subsection{Ethical Considerations and Bias Mitigation} 
As recommender systems increasingly influence what information and products users are exposed to, ethical considerations become paramount. Future research must prioritize the development of mechanisms to identify and mitigate the biases in review-based recommender systems. This includes biases introduced by demographic factors, manipulation through fake reviews, or systemic biases inherent in the data collection process. Ensuring fairness, diversity, and representativeness in recommendations is essential for building equitable systems that serve a broad user base effectively. Review-based recommender systems can benefit from including transparency, ethics, user trust, and data security for fairness and reliability. 

\subsection{Real-time Adaptation and Personalization} 
User preferences are not static; they evolve based on trends, seasons, and personal circumstances. Future review-based recommender systems should aim for real-time adaptation capabilities, where recommendations can dynamically adjust based on emerging trends in review data or changes in user behavior. This calls for the development of agile models that can update themselves frequently and efficiently, possibly through incremental learning or online learning approaches. 

\subsection{Cross-Domain Recommendations}
Exploring ways to apply knowledge from one product area to another and creating a common system to understand user likes and item features are key steps to make recommendation systems better and more flexible across different types of products. Domain adaptation techniques help by using what the system has learned in one area to improve recommendations in another, finding similarities between them \cite{khan2017cross}. At the same time, creating a single framework that captures what users like and item details across all areas can help make recommendations more consistent and relevant, no matter the product type. These methods could lead to better recommendations across various product categories, making suggestions more personalized and useful. 

\subsection{Large Language Models and Recommendations}
LLMs can play a crucial role in generating recommendations by leveraging their natural language understanding capabilities to analyze customer reviews and generate insightful, relevant suggestions \cite{fan2023recommender,laskar2023systematic,laskar2023can}. Recent studies have explored the potential of LLMs in recommender systems, with a focus on prompting engineering \cite{xu2024prompting,he2023large}. These studies have highlighted the effectiveness of LLMs in recommendation tasks, particularly in understanding context, learning user preferences, and generating relevant recommendations. In developing review-based recommendations, LLMs can facilitate sentiment analysis, key theme extraction, personalized recommendations, predictive analysis, and review summarization through methods like zero-shot learning \cite{kojima2022large}, demonstrations \cite{dang2022prompt}, or fine-tuning \cite{raza2023improving}. Additionally, LLMs can improve the processing of unstructured data, such as handling informal language, emojis, and unique formatting used in reviews, which traditional models may struggle to interpret accurately. By handling these complexities more easily, LLMs can provide a more accurate understanding of customer feedback and sentiment, enhancing the overall recommendation quality \cite{liu2023llmrec}. This strategy capitalizes on the advanced natural language understanding capabilities of LLMs, offering customers insightful, targeted recommendations that enhance decision-making and satisfaction \cite{liu2023llmrec}. 

\subsection{Responsible Recommendations}
Most research on recommender systems has traditionally aimed at making the recommended results more accurate. However, the importance of ethical considerations and the impact of these systems on society have increasingly come into focus \cite{wang2022trustworthy}. There is a need for focusing on the aspects of privacy, diversity, and fairness, and how they interconnect with the notion of responsibility in aspect-based recommender systems \cite{elahi2022towards}. While the current literature often lacks a systematic exploration of these integrated approaches, particularly in the context of review-based recommender systems, these systems could greatly benefit from such responsible practices.

%% file: Acknowledgement.tex
\section{Acknowlegements}
This work is supported by the Natural Sciences and Engineering Research Council of Canada and the York Research Chairs program. Jimmy Huang (jhuang@yorku.ca) and Chen Ding (cding@torontomu.ca) are the contact authors of this paper. We would like to extend our gratitude to Syed Faizan for his invaluable assistance in gathering and organizing several papers. We sincerely thank the associate editor and all the anonymous reviewers for their comprehensive feedback and thoughtful insights.